\documentclass[letterpaper, 10 pt, conference]{ieeeconf} 

\IEEEoverridecommandlockouts                              % This command is only needed if 
\usepackage{cite}
\usepackage{amsmath,amssymb,amsfonts}
\usepackage{graphicx} \graphicspath{ {figures/} }
\usepackage{microtype}
\usepackage{makecell}
\usepackage{multirow}
\usepackage{comment}
\usepackage{hhline}
\usepackage{arydshln}
\usepackage{subfig}
\usepackage{balance}
\usepackage{booktabs}
\usepackage{tikz}
\def\checkmark{\tikz\fill[scale=0.4](0,.35) -- (.25,0) -- (1,.7) -- (.25,.15) -- cycle;}
\usepackage[hidelinks]{hyperref}
    
\begin{document}
\title{ \bf
CAT-Net: A Cross-Slice Attention Transformer Model for Prostate Zonal Segmentation in MRI
}
\author{Alex Ling Yu Hung, Haoxin Zheng, Qi Miao, Steven S.~Raman, Demetri Terzopoulos, 
and Kyunghyun Sung
\thanks{Submitted to IEEE Transaction on Medical Imaging on 03/25/2022 for review. This work was supported in part by the National Institutes of Health R01-CA248506 and with funding from the Integrated Diagnostics Program, Departments of Radiological Sciences and Pathology, David Geffen School of Medicine, UCLA}
\thanks{Alex Ling Yu Hung and Haoxin Zheng are with the Department of Computer Science and the Department of Radiological Sciences, University of California, Los Angeles (UCLA), CA 90095, USA (e-mail: alexhung96@ucla.edu, haoxinzheng@ucla.edu).}
\thanks{Qi Miao is with the Department of Radiological Sciences, University of California, Los Angeles (UCLA), CA 90095, USA, and the Department of Radiology, The First Affiliated Hospital of China Medical University, Liaoning Province 110001, China (e-mail: meganmiaoqi@126.com).}
\thanks{Steven S Raman and Kyunghyun Sung is with the Department of Radiological Sciences, University of California, Los Angeles (UCLA), CA 90095 USA (e-mail: SRaman@mednet.ucla.edu, ksung@mednet.ucla.edu).}
\thanks{Demetri Terzopoulos is with the Department of Computer Science, University of California, Los Angeles (UCLA), CA 90095, USA, and is a co-founder of VoxelCloud, Inc., Los Angeles, CA 90024, USA (e-mail: dt@cs.ucla.edu)}}
\maketitle
\thispagestyle{empty}
\pagestyle{empty}

\begin{abstract}
Prostate cancer is the second leading cause of cancer death among men in the United States. The diagnosis of prostate MRI often relies on accurate prostate zonal segmentation. However, state-of-the-art automatic segmentation methods often fail to produce well-contained volumetric segmentation of the prostate zones since certain slices of prostate MRI, such as base and apex slices, are harder to segment than other slices. This difficulty can be overcome by leveraging important multi-scale image-based information from adjacent slices, but current methods do not fully learn and exploit such cross-slice information. In this paper, we propose a novel cross-slice attention mechanism, which we use in a Transformer module to systematically learn cross-slice information at multiple scales. The module can be utilized in any existing deep-learning-based segmentation framework with skip connections. Experiments show that our cross-slice attention is able to capture cross-slice information significant for prostate zonal segmentation in order to improve the performance of current state-of-the-art methods. Our method improves segmentation accuracy in the peripheral zone, such that the segmentation results are consistent across all the prostate slices (apex, mid-gland, and base). 
\end{abstract}

\section{Introduction}

Prostate cancer (PCa) is the most common cancer and the second leading cause of cancer-related death among men in the United States~\cite{rawla2019epidemiology}. Multi-parametric MRI (mpMRI), including T2-weighted imaging (T2WI), diffusion-weighted imaging (DWI), and dynamic contrast-enhanced (DCE) MRI, 
is now the preferred non-invasive imaging technique for prostate cancer (PCa) diagnosis prior to biopsy~\cite{appayya2018national}. %Prostate Imaging Reporting and Data System version 2.1 (PI-RADS v2.1), the current clinical standard for interpreting mpMRI in the diagnosis of PCa, suggests that different contrast of MR images should play different roles when diagnosing lesions in different areas of the prostate~\cite{turkbey2019prostate}. 
According to the Prostate Imaging Reporting and Data System (PI-RADS v2.1)~\cite{turkbey2019prostate}, the current clinical standard for interpreting mpMRI, a suspicious lesion should be analyzed differently in different prostate zones, among them the transition zone (TZ) and the peripheral zone (PZ), due to variations in image appearance and cancer prevalence~\cite{israel2020multiparametric}.
The zonal information is essential and should be provided explicitly for the accurate identification and assessment of suspicious lesions. Moreover, the size of the TZ is often used to evaluate and monitor benign prostate hyperplasia (BPH) in clinical practice~\cite{lawrentschuk2016benign}. However, the manual annotation of prostate zones is typically time-consuming and highly variable depending on experience level. Therefore, reliable and robust automatic zonal segmentation methods are needed in order to improve PCa detection.

With the emergence of deep learning (DL), multiple DL-based medical image segmentation methods have been proposed to automatically segment the prostate zones~\cite{bardis2021segmentation,ronneberger2015u,liu2020exploring}. Several studies~\cite{liu2019automatic,nai2020evaluation} report that particular parts of the prostate, such as the apex and base locations, are more difficult to segment than the mid-gland slices due to high segmentation uncertainties and ambiguity on the zonal boundaries, which can be improved by considering information from nearby slices. However, most 2D-based segmentation methods do not fully consider or systematically learn the available cross-slice information, thus they may disregard important structural information about the prostate, leading to inconsistent segmentation results. It is crucial to take the through-plane information or cross-slice relationship into full consideration when devising prostate zonal segmentation models.

Several 3D DL-based medical image segmentation networks have been proposed in previous studies~\cite{cciccek20163d,yu2017automatic}, but their application to prostate MRI has been limited. Following the standard guideline of PI-RADS~\cite{turkbey2019prostate}, T2WI images are acquired using the multi-slice 2D Turbo Spin Echo (TSE) sequence, resulting in high in-plane image resolution (e.g., 0.3--1.0\,mm) but low through-plane resolution (e.g., 3.0--6.0\,mm). Due to the anisotropic nature of the image resolution, existing 3D segmentation networks may not be directly applicable as they are typically designed for nearly isotropic 3D images~\cite{cciccek20163d,yu2017automatic}, and the performance would not be ideal when anistropic data were used~\cite{jia20193d,isensee2018nnu,isensee2017automatic}. 

Transformer networks have become the dominant DL architecture in the natural language processing (NLP) domain as their multi-head self-attention (MHSA) mechanisms learn the global context and the relationship among different word embeddings~\cite{vaswani2017attention}. Following similar ideas, MHSA mechanisms have recently been applied in the computer vision domain to address tasks such as object detection, segmentation, and classification.

\begin{figure}[h!]
     \centering
     \subfloat[]{
         \includegraphics[width=0.45\textwidth]{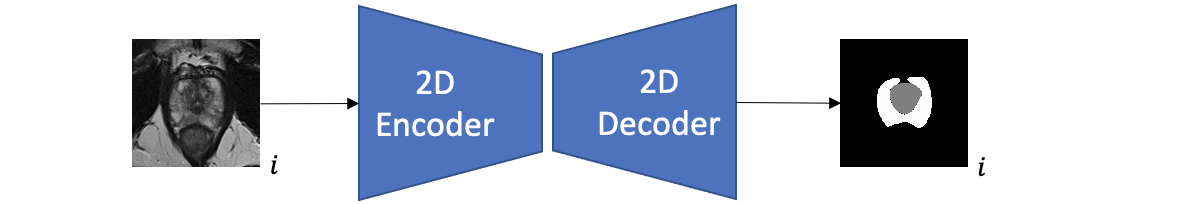}
         \label{2dnet}
         }\\
         \subfloat[]{
         \includegraphics[width=0.45\textwidth]{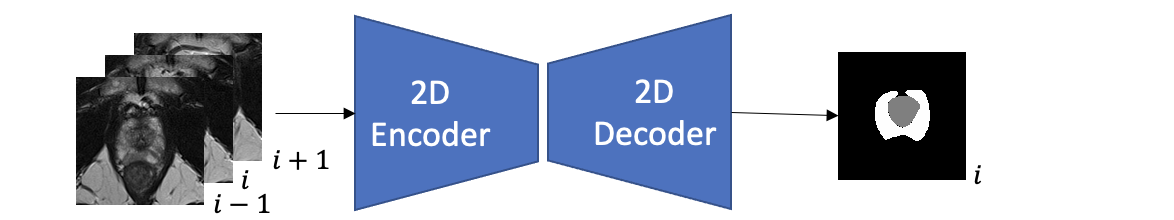}
         \label{25dnet}
         }\\
         \subfloat[]{
         \includegraphics[width=0.45\textwidth]{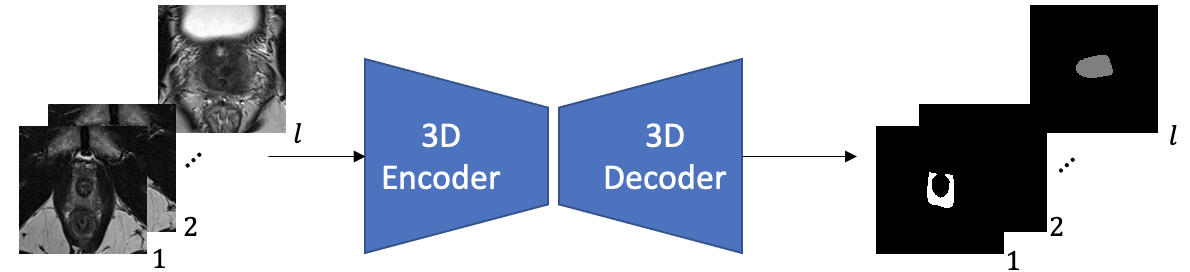}
         \label{3dnet}
         }\\
         \subfloat[]{
         \includegraphics[width=0.45\textwidth]{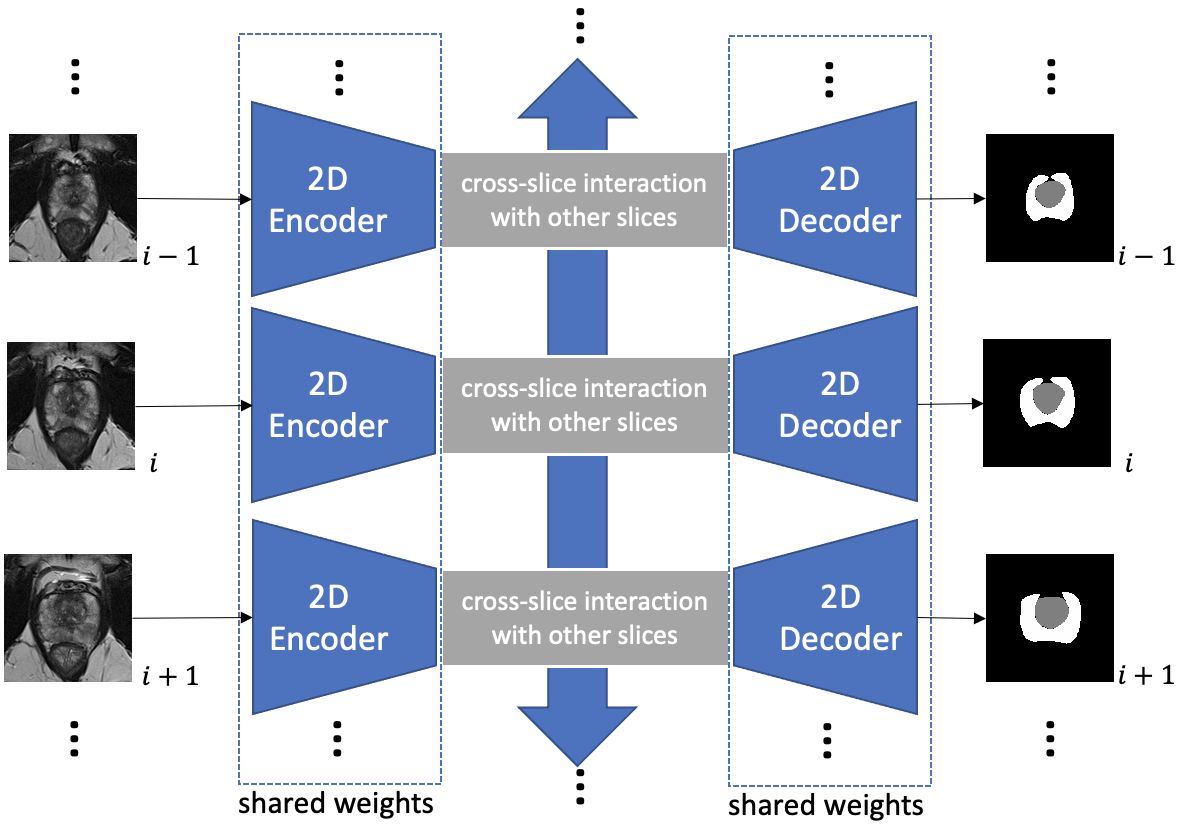}
         \label{ournet}
         }\\
     \caption{(a) A conventional 2D network, which takes in the slice of interest and predicts the segmentation mask of that slice. (b) An example of conventional 2.5D network that takes in 2 nearby slices, defined as 2.5D w/3. It takes in slice $i-1$, $i$, $i+1$, and only predicts the segmentation of the middle slice which is slice $i$. (c) A conventional 3D network, where it takes in the entire volume with $l$ slices and performs 3D convolutions. It outputs the segmentation masks for all the $l$ slices. (d) A simple illustration of our proposed framework, where we only use 2D convolutions. After encoding all the slices in the same volume, the encoded feature maps from different slices go through the attention-based cross-slice interaction, then go back into the 2D decoder.}
     \label{dif_nets}
 \end{figure}

As it can capture the long-range dependencies in images~\cite{vaswani2017attention,valanarasu2021medical,tang2018self}, self-attention is a promising mechanism for systematically learning the cross-slice information needed to improve 3D medical image analysis; that is, as compared to convolutional networks whose inherent inductive bias is to focus more on neighboring features. To exploit the fact that accurate prostate zonal segmentation can benefit from information spanning all the slices through the entire prostate instead of only neighboring slices, we devise a 2.5D cross-slice attention-based module that can be incorporated within any network architecture with skip connections in order to capture the global cross-slice relationship. Note that a 2.5D method employs only 2D convolutional layers yet learns 3D information. In contrast, 2D method also only uses 2D convolutional layers but only considers one single image, while 3D method takes a 3D volume as input and uses 3D convolutional layers. A simple illustration of conventional 2D, 2.5D, 3D methods, and our proposed method is shown in Fig.~\ref{dif_nets}. Our main contributions include the following:
(1) We formally propose the use of an cross-slice attention mechanism to capture the relationship between MRI slices for the purposes of prostate zonal segmentation.
(2) We devise a novel 2.5D Cross-slice Attention Transformer (CAT) module that can be incorporated into any existing skip-connection-based network architecture to exploit long-range information from other slices.
(3) We perform an experimental study which demonstrates that our proposed DL models, called CAT-Net, is able to improve the zonal segmentation results both in general and on different prostate parts, resulting in a new state of the art performance.

\section{Related Work}

%In natural image domain, Long et al.~\cite{long2015fully} introduced the use of a fully convolutional network (FCN) to perform image segmentation as a pixel-wise classification task. DeepLab~\cite{chen2014semantic} then became the dominate backbone for natural image segmentation, which uses dialated convolutions to enlarge the FCN's field of view and a fully-connected conditional random field (CRF) for post processing. On top of DeepLab, DeepLabV2\cite{chen2017deeplab} came up with atrous spatial pyramid pool (ASPP) to account for different object scales. DeepLabV3~\cite{chen2017rethinking} and DeepLabV3+~\cite{chen2018encoder} expanded the idea of ASPP and achieve better segmentation results on natural images. 

%In medical domain, 
U-Net~\cite{ronneberger2015u} has revolutionized medical image segmentation using an encoder-decoder convolutional neural network (CNN) architecture with multi-scale skip connections between the encoder and decoder that preserve high-resolution image information. Building upon U-Net, a number of subsequent segmentation models have been proposed. ResU-Net~\cite{khanna2020deep} applied the residual connections from ResNet~\cite{he2016deep}, and the combination of ResNet with U-Net has also proven effective outside of medical image segmentation~\cite{zhang2018road,diakogiannis2020resunet}. Based on ResU-Net, Alom et al.~\cite{alom2019recurrent} proposed a segmentation network with better feature representation by means of feature accumulation with recurrent convolutional layers. Apart from residual blocks, other attention-based modules have been incorporated into U-Net to improve segmentation; e.g., Rundo et al.~\cite{rundo2019use} incorporated a squeeze and excitation (SE) module~\cite{hu2018squeeze}, a form of channel-wise attention, and Oktay et al.~\cite{oktay2018attention} took the attention module~\cite{jetley2018learn} a step further and applied it in U-Net. Other U-Net variants have recently emerged. nnU-Net~\cite{isensee2018nnu} modified the batch normalization~\cite{ioffe2015batch} in U-Net to instance normalization~\cite{ulyanov2016instance} and employed a leaky rectified linear activation unit (ReLU). Compared to U-Net, U-Net++~\cite{zhou2018unet++} used more nested and dense skip connections to better capture the fine-grained details of foreground objects. MSU-Net~\cite{su2021msu} added multi-scale blocks, which consist of convolutions with different kernel sizes, into U-Net to improve the segmentation details. Gu et al.~\cite{gu2019net} proposed a novel dense atrous convolution (DAC) block along with a residual multi-kernel pooling (RMP) block and put them in the bottleneck layer of U-Net to capture more high-level features and preserve more spatial information. Apart from U-Net based network architectures, DRINet~\cite{chen2018drinet} consists of dense connection blocks, residual Inception blocks, and unpooling blocks, which can learn more distinctive features. Crossbar-Net~\cite{yu2019crossbar} samples vertical and horizonal patches and processes them separately on two sub-models. 

Bardis et al.~\cite{bardis2021segmentation} applied the U-Net on T2WI images to perform prostate zonal segmentation. Building on top of DeepLabV3+~\cite{chen2018encoder}, Liu et al.~\cite{liu2019automatic} employed multi-scale feature pyramid attention (MFPA) to further utilize encoder-side information at different scales to segment prostate zones, and subsequently expanded the network structure with a spatial attentive module (SAM) and Bayesian epistemic uncertainty based on dropout~\cite{liu2020exploring}.
%\cite{kendall2017uncertainties}. 
Cuocolo et al.~\cite{cuocolo2021deep} have thoroughly investigated network structures for prostate zonal segmentation and concluded that ENet~\cite{paszke2016enet} is superior to U-Net and ERFNet~\cite{romera2017erfnet}. Zabihollahy et al.~\cite{zabihollahy2019automated} proposed to use two separate networks for the segmentation of different zones and combine them with post-processing. Not limited to T2WI images, Rundo et al.~\cite{rundo2018fully} performed multispectral MRI prostate gland segmentation based on clustering.

3-dimensional CNNs are popular for 3D medical image segmentation. 3D U-Net~\cite{cciccek20163d}, VNet~\cite{milletari2016v}, and DenseVoxelNet~\cite{yu2017automatic} are U-Net architectures that use 3D rather than 2D convolution, and they have proven effective on image data whose cross-pixel distance is similar in all three dimensions. Wang et al.~\cite{wang2018two} used a two-stage 3D U-Net for multi-modality whole heart segmentation. Other researchers have applied modified 3D U-Nets to infant brain segmentation \cite{qamar2020variant}, lung nodule segmentation \cite{xiao2020segmentation}, and brain tumor segmentation \cite{chang2018brain}. In terms of the application of 3D methods on prostate segmentation, Nai et al.~\cite{nai2020evaluation} investigated how different 3D algorithms would work on prostate zonal segmentation, and concluded that most methods have similar performance but the addition of ADC and DWI data would slightly improve the performance. Yu et al.~\cite{yu2017volumetric} used mixed residual connections in a volumetric ConvNet for whole prostate segmentation. Z-Net~\cite{zhang2019prostate}, capable of capturing more features in a multi-level fashion, was built on top of U-Net to perform 3D prostate segmentation. Wang et al.~\cite{wang2019deeply} incorporated group dilated convolution into a deeply supervised fully convolutional framework for prostate segmentation.

Originally developed for NLP, the Transformer architecture~\cite{vaswani2017attention} has recently been gaining momentum in vision. Dosvitskiy et al.~\cite{dosovitskiy2020image} proposed the Vision Transformer (ViT) for image classification, treating images as 16$\times$16 words. Unlike the ViT, the Swin Transformer~\cite{liu2021swin} used shifted windows instead of $16\times16$ fixed-size windows. Carion et al.~\cite{carion2020end} proposed the Detection Transformer (DETR) for object detection. 

Several Transformer-based methods have been proposed for medical image segmentation. MedT~\cite{valanarasu2021medical} couples a gated position-sensitive axial attention mechanism with a Local-Global (LoGo) training methodology, improving segmentation quality over the U-Net and attention U-Net. UTNet~\cite{gao2021utnet} incorporates Transformer blocks into the skip connections in U-Net, allowing the skip connection feature maps to go through a Transformer Block before they reach the decoder side. CoTr~\cite{xie2021cotr} instead concatenates all the skip connection feature maps and applies the Transformer on the concatenated vector. Petit et al.~\cite{petit2021u} proposed a U-Net based Transformer framework with a self attention or cross attention module after each skip connection of the normal U-Net. However, these Transformer-based medical image segmentation methods apply attention only between pixels or patches, not between slices. 

Our work is closest to that by Guo and Terzopoulos~\cite{guo2021transformer}, which utilized attention between slices at the bottom layer of a U-Net with a Transformer network. The work demonstrated its feasibility for 3D medical image segmentation but did not consider cross-slice information at different scales and different semantic information learned by multiple heads of the attention mechanism. 

\section{Methods}
\label{methods}

\subsection{Overview}

\begin{figure*}
     \centering{
         \includegraphics[width=0.98\textwidth]{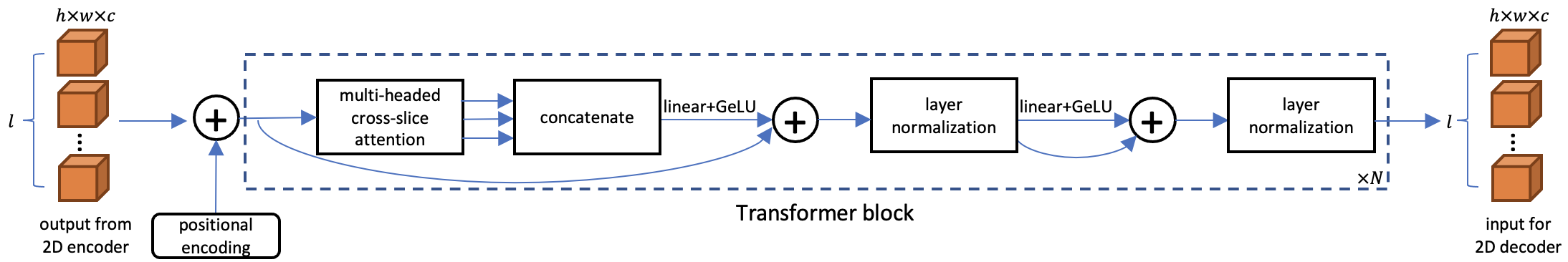}
         }
     \caption{The CAT module includes a positional encoding and $N$ Transformer blocks.}
     \label{ISIM}
 \end{figure*}

\begin{figure*}
     \centering
     \subfloat[]{
         \includegraphics[height=0.325\textwidth]{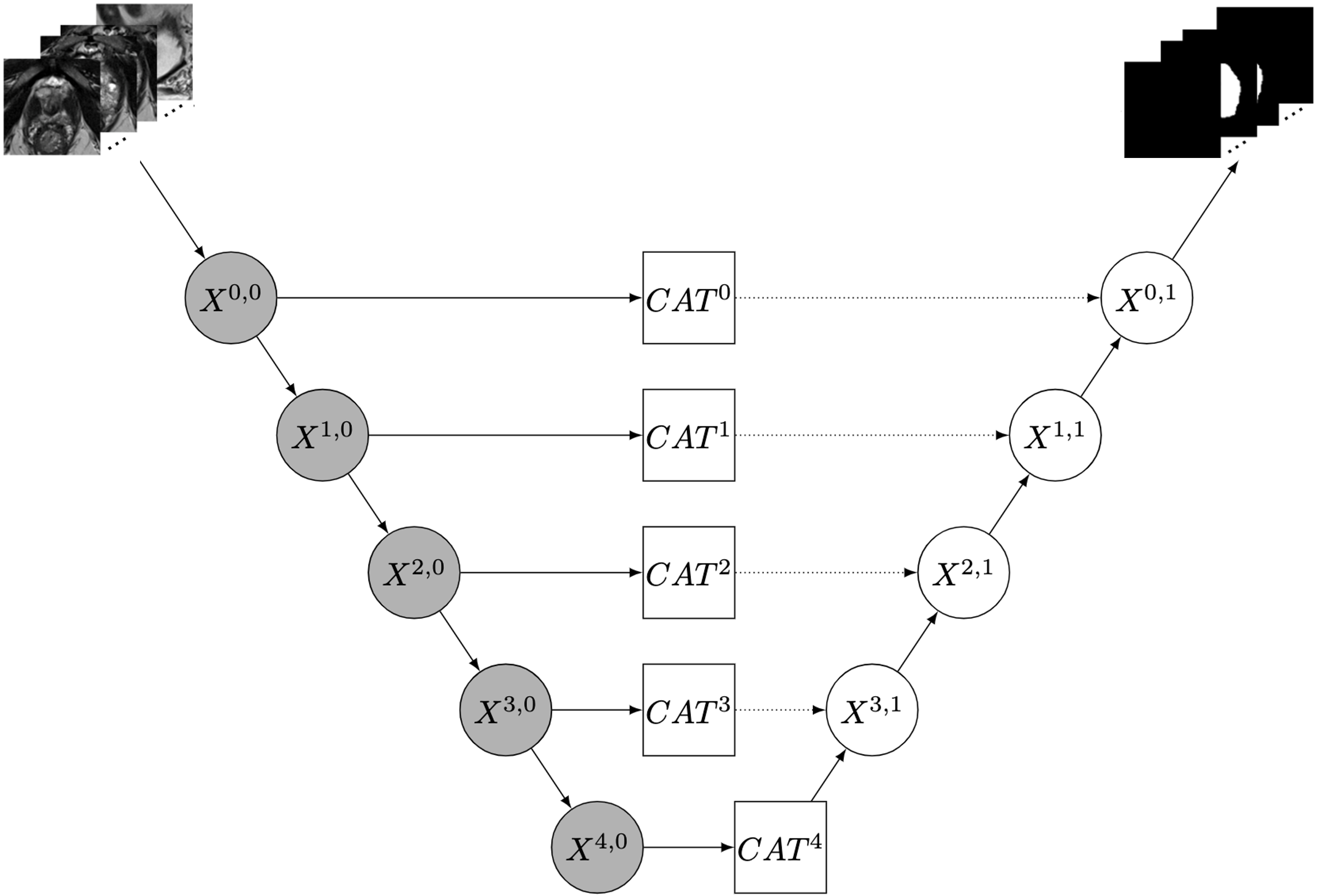}
         \label{nnunet}
         }
         \subfloat[]{
         \includegraphics[height=0.325\textwidth]{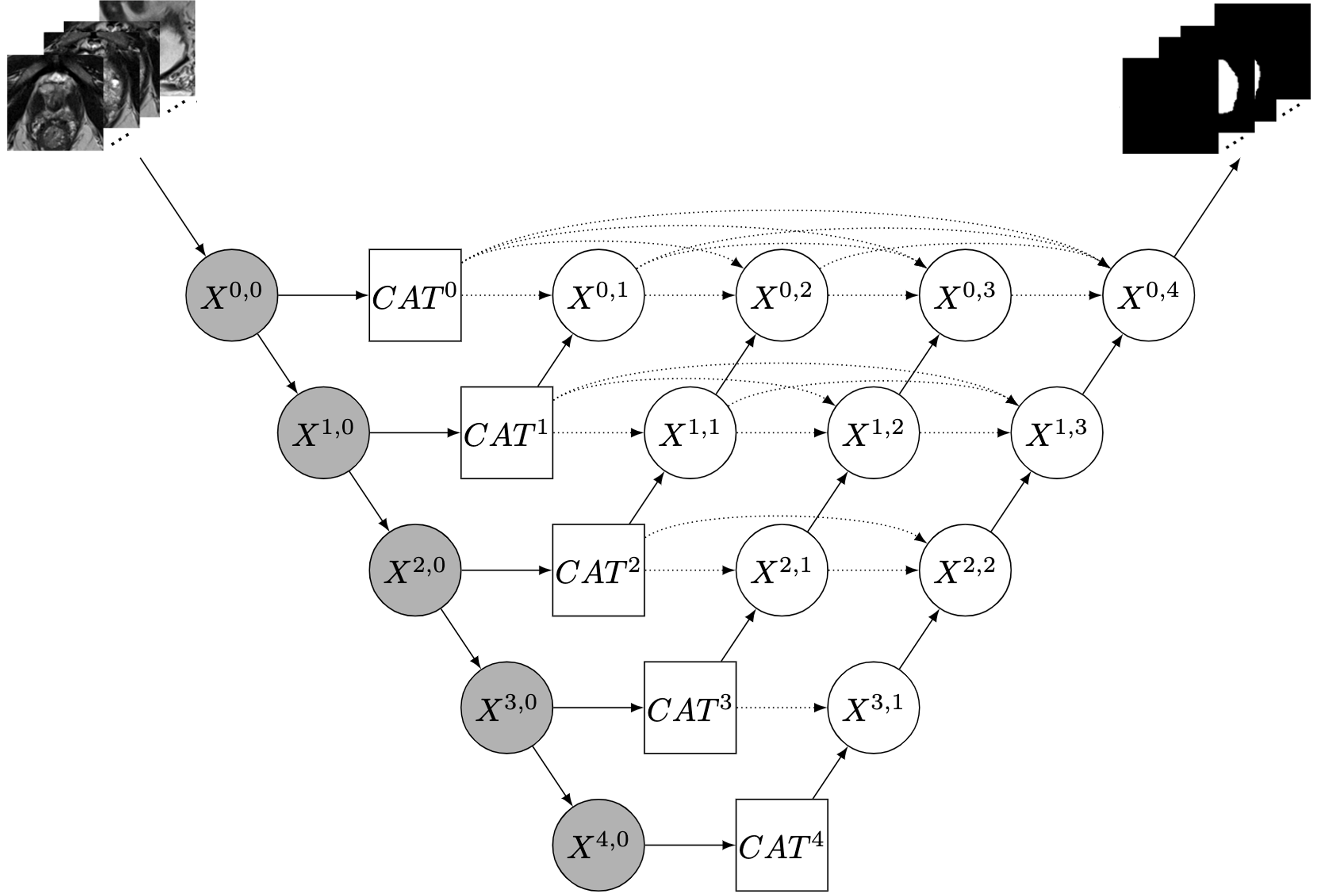}
         \label{unet++}
         }
     \caption{The implementation of CAT modules in the (a) nnU-Net and (b) nnU-Net++. Following the notation in~\cite{zhou2018unet++}, circles represent feature maps calculated at the corresponding node; gray circles represent the 2D encoder whereas white circles represent the 2D decoder. Rectangles represent the CAT modules, which operate in 3D. The input stack of images from the patient flow through the encoder, into CAT modules, and then through the decoder to produce the output segmentation.}
     \label{network}
 \end{figure*}

We propose a 2.5D Cross-Slice Attention Transformer (CAT) module to systematically learn cross-slice information for prostate zonal segmentation. The module is applicable within U-Net-like architectures with skip connections between the encoder and decoder. Segmentation models using CAT modules consist of three parts: a standard 2D encoder, a standard 2D decoder, and CAT modules in different layers. In other words, the network would be a pure 2D network if we remove the CAT modules, where there would be no interaction between slices. The overall structure of the CAT module is illustrated in Fig.~\ref{ISIM}, and the incorporation of CAT modules into existing deep models, such as nnU-Net and nnU-Net++, is illustrated in Fig.~\ref{network}. 

The remainder of this section is organized as follows: Section~\ref{intersliceattention} introduces the cross-slice attention mechanism. Cross-slice attention is used in the Transformer block, which is discussed in Section~\ref{transformerblock}. Positional encoding followed by $N$ Transformer blocks comprise the CAT module, which is described in Section~\ref{networkarchitecture}, where we also explain how the CAT module is used in existing networks. In Section~\ref{networkarchitecture}, we mainly discuss how the CAT module can be incorporated into U-Net and U-Net++ networks to yield our novel CAT-Net models, but any other skip-connection-based networks are also amenable.
 
\subsection{Cross-Slice Attention}
\label{intersliceattention}

Previous studies have used attention modules to learn inter-channel or inter-pixel relationships, whereas our goal here is to use the attention mechanism to learn the cross-slice relationship for the purposes of our segmentation task. 
This is important because single-slice prostate zonal segmentation can suffer from ambiguities, especially near apex and base slices. 
This is also true for manual annotation as clinicians typically refer to nearby slices while annotating the current slice of interest. 
We devise an algorithm that mimics the manual segmentation process by attending to other slices when the current slice is being annotated. To this end, we regard each slice as being analogous to a word in NLP problems while keeping the spatial information of images intact. After the images are encoded by the encoder, we treat the image features as a deep representation of the ``word''.

Let an $l$-slice stack of input images be represented by the 4D tensor $x\in\mathbb{R}^{l\times h\times w\times c}$, which is a stack of feature maps of height $h$, width $w$, and number of channels $c$. 
To mitigate computational expense, we leverage the typically high correlation of nearby pixels in feature maps and downsample $x$ by average pooling with a kernel size of $k$, generating a condensed stack of feature maps $x_\text{pool}\in\mathbb{R}^{l\times\frac{h}{k}\times\frac{w}{k}\times c}$. 
We then calculate queries $Q'$, keys $K'$ as a linear projection of $x_\text{pool}$, and values $V$ as a linear projection of $x$ in the channel dimension:
\begin{align}
    Q'&=x_\text{pool}W_Q,\\
    K'&=x_\text{pool}W_K,\\
    V&=xW_V,
\end{align}
where $W_Q,W_K,W_V\in\mathbb{R}^{c\times c}$ are learnable weights. Note that $Q'$, $K'$, and $V$ are the deep representation of the $l$ slices and that the above 4D tensor multiplication is defined as follows: The product $C=BW$ of a $l\times h\times w\times c$ tensor $B$ and a $c\times c$ matrix $W$ is calculated as
\begin{equation}
    C[i,j,k,m]=\sum_{n=1}^c B[i,j,k,n]W[n,m].
\end{equation}

The attention matrix $A\in\mathbb{R}^{l\times l}$, which represents how much attention the algorithm should pay to other slices while segmenting a slice, is calculated as
\begin{equation}
    A=\text{softmax}\left(\frac{QK^T}{\sqrt{hwc/k^2}}\right),
\end{equation}
where each row of $Q,K\in\mathbb{R}^{l\times\frac{hwc}{k^2}}$, the reshaped matrices of $Q'$ and $K'$, represents the query and key for each slice. The softmax is performed on the second dimension. An element $A[i,j]$ in the attention matrix indicates how similar the query for slice $i$ is to the keys for slice $j$; i.e., it is computed as a function that performs a weighted average of the values for all the slices to account for the interactions between queries and keys. 
The output of the cross-slice attention $y\in\mathbb{R}^{l\times h\times w\times c}$ is computed as
\begin{equation}
    y=AV.
\end{equation}

\begin{figure*}
     \centering
     \includegraphics[width=0.98\textwidth]{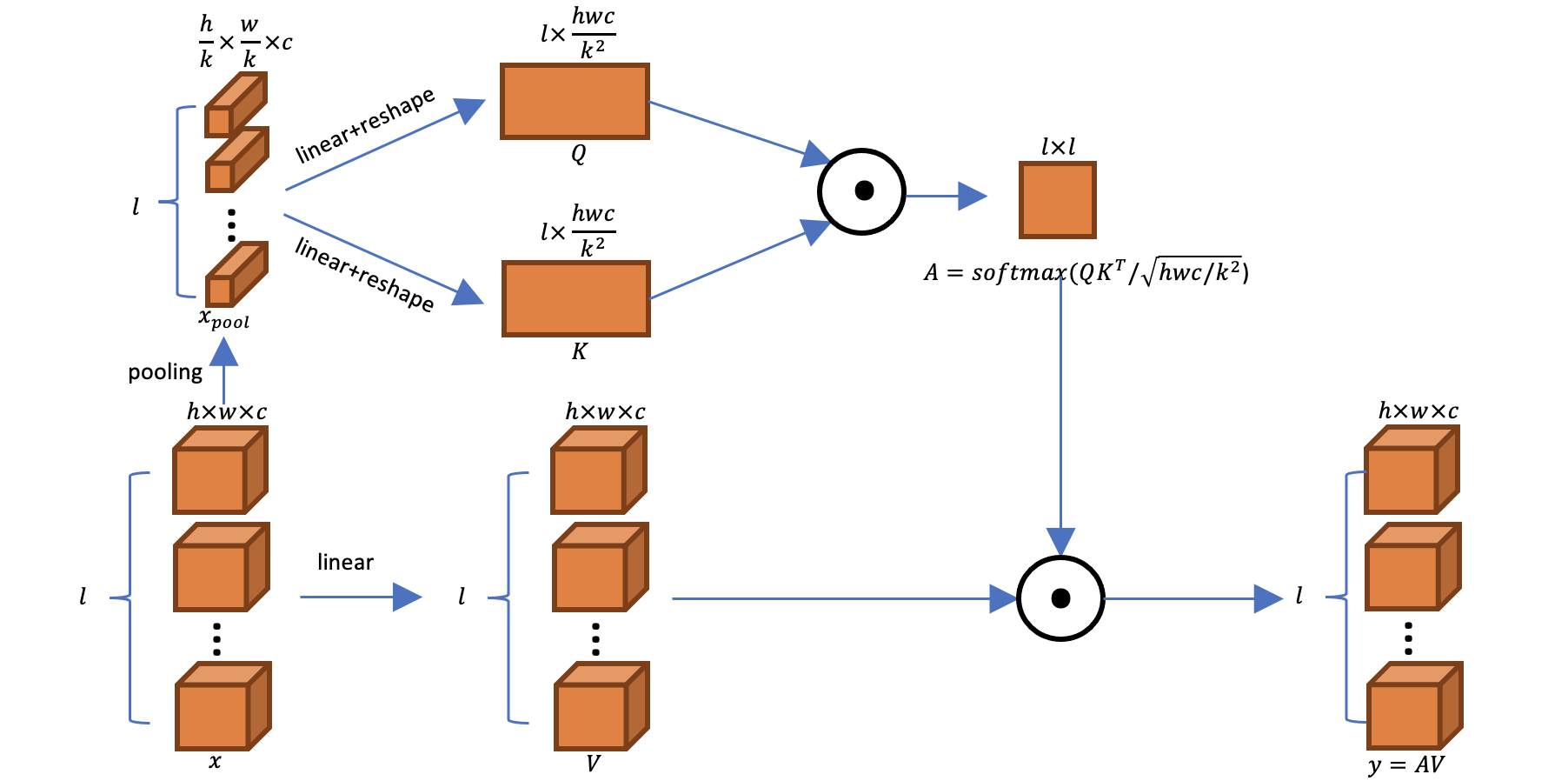}
     \caption{Cross-slice attention with input $x$ and output $y$.}
     \label{attentionn}
 \end{figure*}
 
Fig.~\ref{attentionn} illustrates our cross-slice attention mechanism, which, unlike the 2D self-attention mechanism~\cite{vaswani2017attention}, uses a condensed deep image feature map to calculate queries $Q$ and keys $K$ and uses a normally encoded feature map for values $V$, while working in a 4D space to calculate the attention matrix between slices instead of between pixels. 

\subsection{Transformer Block}
\label{transformerblock}

We incorporate our cross-slice attention mechanism into a Transformer block, which is the structure widely used in Transformer-based approaches, where one typically sees a multi-headed attention module followed by some linear operations, non-linear modules, and normalizations with some skip connections in between. Our Transformer block is shown within the dashed box in Fig.~\ref{ISIM}. The input to the Transformer block goes through a multi-headed cross-slice attention and is then subjected to some linear projections along with some non-linear activations. Specifically, we perform multiple cross-slice attention in parallel, obtaining $y_i$, where $i=1,2,\dots,H$ ($H$ is the number of heads), similar to previous work~\cite{vaswani2017attention}. This is done so that the network can learn multiple semantics during the attention procedure; i.e., for different meanings, the network learns different attention matrices. Subsequently, we concatenate all the $y_i$ in the $c$ dimension to obtain the output $y$ of the multi-headed cross-slice attention module. The final output of the Transformer block can be expressed as
\begin{equation}
    z=\text{Layer\_Norm}(\text{GELU}(z_\text{int}W_2+b_2)+z_\text{int}),
\end{equation}
with intermediate result
\begin{equation}
    z_\text{int}=\text{Layer\_Norm}(\text{GELU}(yW_1+b_1)+x),
\end{equation}
where $W_1$ and $W_2$ are the linear projection matrices, $b_1$ and $b_2$ are bias terms, $\text{GELU}$ is the Gaussian error linear unit~\cite{hendrycks2016gaussian}, and $\text{Layer\_Norm}$ performs layer normalization~\cite{ba2016layer} across the $h$, $w$, and $c$ dimensions.

\subsection{Network Architecture}
\label{networkarchitecture}

As shown in Fig.~\ref{network}, we use nnU-Net~\cite{isensee2018nnu} and nnU-Net++~\cite{zhou2018unet++} as the backbones of our CAT-Net architectures, which we denote as CAT-nnU-Net and CAT-nnU-Net++, respectively. Other network architectures with skip connections between the encoder and decoder would also be suitable. We find that our attention mechanism works well on nnU-Net-like designs where Leaky ReLU and instance normalization is used rather than normal ReLU and batch normalization. 

The 2D encoder $E$ takes in a $l\times c_0\times h_0\times w_0$ tensor, where $l$, $h_0$, $w_0$, and $c_0$ are the number of slices, height, width and the number of channels, respectively. It treats $l$ as the batch dimension, where the slices do not interfere with each other. A deep feature map $x_i\in\mathbb{R}^{l\times h_i\times w_i\times c_i}$ is input into the CAT module (Fig.~\ref{ISIM}), in which it is subjected to a positional encoding and $N$ Transformer blocks. The positional encoding is important in this task, since it tells the network the location of the slice. As in previous work~\cite{vaswani2017attention}, we first add to it a learnable positional encoding $\textit{PE}_i$ initialized as follows~\cite{gehring2017convolutional}: For all elements on slice $p$ and channel $2j$, 
\begin{equation}
    \textit{PE}_i[p,:,:,2j]=\sin\left(\frac{p}{10000^{2j/c_i}}\right),
\end{equation}
and for all the elements on slice $p$ and channel $2j+1$,
\begin{equation}
    \textit{PE}_i[p,:,:,2j+1]=\cos\left(\frac{p}{10000^{2j/c_i}}\right).
\end{equation}
Then, the feature is passed through $N$ Transformer blocks, where the feature maps of different slices interact to yield the outputs of the skip connection $z_i\in\mathbb{R}^{l\times h_i\times w_i\times c_i}$. Again, we treat the first dimension as the batch dimension, interpreting $z_i$ as $l$ feature maps of 2D images. The 2D decoder takes in $z_i$ and outputs the final segmentation masks.

More generally, encoder $E$ is given $l$ image slices denoted as $x_0\in\mathbb{R}^{l\times h_0\times w_0\times c_0}$ and returns the encoded images $x_i\in\mathbb{R}^{l\times h_i\times w_i\times c_i}$  at different scales $1\leq i\leq L$, where $L$ is the total number of layers in the encoder: $\{x_i\}_{1\leq i\leq L}=E(x_0)$.

We denote the lowest layer in the decoder as $D_L$ and the layers above it as $D_{L-1}, D_{L-2}, \dots$, and the output of decoder layer $i$ as $d_i$. In a conventional skip-connection-based network, the decoder takes in the feature maps from different scales:
\begin{equation}
    d_i=\left\{ 
  \begin{array}{ l l }
    D_i(x_i) & \quad \textrm{if } i=L,\\
    D_i(x_i,d_{i+1})              & \quad \textrm{otherwise,}
  \end{array}
\right.
\end{equation}
whereas in our CAT-Net, CAT modules are inserted into the network:
\begin{equation}
    d_i=\left\{ 
  \begin{array}{ l l }
    D_i(\text{CAT}_i(x_i)) & \quad \textrm{if } i=L,\\
    D_i(\text{CAT}_i(x_i),d_{i+1})              & \quad \textrm{otherwise,}
  \end{array}
\right.
\end{equation}
where $\text{CAT}_i$ is the CAT module at scale $i$. 

Our method works better on networks with skip connections at different scales since multi-resolution feature maps capture different semantic information~\cite{gatys2015neural,olah2017feature}; thus, the attention at different scales should not be the same. The cross-slice attention matrices should be different across the different layers of the network, whereas in networks without skip connections across the different scales, the semantic information at different scales is not represented in the feature maps, and the same attention matrix would be applied, leading to unsatisfactory results.

\section{Experiments}

Our experimental study was performed in compliance with the United States Health Insurance Portability and Accountability Act (HIPAA) of 1996 and was approved by the institutional review board (IRB) with a waiver of the requirement for informed consent. 

\subsection{Study Population and MRI Data}
\subsubsection{Internal Dataset}
296 patients who underwent pre-operative 3 Tesla (3T) mpMRI prior to robotic-assisted laparoscopic prostatectomy were included in the study. Patients with prior radiotherapy or hormonal therapy and with an endorectal coil were excluded. All mpMRI scans were performed on one of the four 3T MRI scanners (Trio, Skyra, Prisma, Vida; Siemens Healthineers, Erlangen, Germany) from January 2013 to December 2018 at a single academic institution. T2WI was acquired using a T2-weighted Turbo Spin Echo (TSE) MR sequence following the standardized imaging protocol of the European Society of Urogenital Radiology (ESUR) PI-RADS guidelines~\cite{turkbey2019prostate}. The T2WI images were used for the zonal segmentation with an in-plane resolution of 0.625\,mm$^2$, a through-plane resolution of 3\,mm, and an image size of $320\times320\times20$ voxels. In this study, we cropped the central images to $128\times128$ and use 238, 29, and 29 patients for training, validation, and testing, respectively. For experiments involving cross validation, we randomly mixed up the data and put them into 5 folds. 

\subsubsection{ProstateX Dataset}
A total of 193 patients from the ProstateX \cite{litjens2014computer} dataset were included in the study. The imaging was performed by the MAGNETOM Trio and Skyra Siemens 3T MR scanners. A turbo spin echo sequence was used to acquire T2WI images, which had a resolution of 0.5\,mm in plane and a slice thickness of 3.6\,mm. The image size is $384\times384\times18$ voxels, where we pick only the middle 18 slices for each patient. In this study, we cropped the central images to $160\times160$ and resampled them to $128\times128$ and use 157, 20, and 20 patients for training, validation, and testing, respectively.
\subsection{Implementation Details}

For our nnU-Net implementation, we downsampled 5 times on the encoder side, and had 64, 128, 256, 512, 1024, and 2048 filters in each of the convolutional layers of the encoder. For our nnU-Net++ implementation, we downsampled 4 times on the encoder side and had 64, 128, 256, 512, and 1024 filters in each of the convolutional layers of the encoder. The decoders were the exact opposite of the encoders. For the 3D networks, we downsampled twice and once in the networks for our internal dataset and ProstateX dataset respectively because the each patient has a volume has 20 and 18 slices in each of the dataset. This is one of the limitation of 3D networks, where you might not perform multiple downsampling operations due to the total slice number. For other parts of the 3D networks and all the other models in our experiments, we strictly followed the architectures described in the original papers.

We applied cross entropy loss as the loss function for 150 epochs in all the training procedures, and used Adam~\cite{kingma2014adam} with a learning rate of 0.0001 and weight decay regularization~\cite{loshchilov2018fixing} with the parameter set to $1\times10^{-5}$. In our method, we set the number of Transformer blocks $N=2$, the number of heads $H=3$, and the average pooling size $k=4$. The CAT-Nets were trained on a single Nvidia Quadro RTX 8000 GPU, while the other networks were trained on an Nvidia RTX 3090 GPU.

\subsection{Evaluation}

\subsubsection{Evaluation Metrics} 

We used Intersection over Union (IoU), Dice coefficient (Dice), Relative absolute volume difference (RAVD), and average symmetric surface distance (ASSD) for evaluation. All these evaluation metrics were calculated in a 3D patient-wise manner.

For experiments involving statistical testing, we used Mann-Whitney U Test~\cite{nachar2008mann} to statistically test the distribution of results from our method and the competing methods.

\subsubsection{Quantitative Evaluation}

We performed a comprehensive comparison between our method and state-of-the-art 2D, 2.5D, and 3D medical image segmentation methods.

With regard to 2D methods, we compared ours against DeepLabV3+ ~\cite{chen2018encoder}, Liu et al.'s method~\cite{liu2020exploring}, CE-Net-\cite{gu2019net}, MSU-Net~\cite{su2021msu}, nnU-Net++~\cite{zhou2018unet++}, and nnU-Net~\cite{isensee2018nnu}. To be specific, we adopted the 2D nnU-Net from the paper~\cite{isensee2018nnu}. Besides, we also adapted the Leaky ReLU along with the instance normalization design, which is the main architectural contribution by nnU-Net, into U-Net++ and denote it as nnU-Net++. We did not directly compare against U-Net~\cite{ronneberger2015u} and U-Net++ because previous work has demonstrated the superiority of nnU-Net-based designs, supported by the experimental results in Section~\ref{sec_ablation}. 

We continued using nnU-Net and nnU-Net++ as the backbones of our 2.5D methods. Like Zhang et al.~\cite{zhang2019multiple}, we stacked nearby slices together and input the stacks to 2D networks in order to segment the middle slices. In other words, during each run, the networks take in a stack of images and only outputs the segmentation mask for the middle slice rather than the masks for all the slices in the stack. For example, if a stack of slice $i-1$, $i$ and $i+1$ is fed into the network, then the network would output the segmentation of slice $i$. In our experiments, we stacked 3, 5, and 7 slices together and fed them into the networks, which we denote as nnU-Net and nnU-Net++ w/3, w/5, and w/7. Note that nnU-Net w/1 and nnU-Net++ w/1 are the 2D networks where we only input 1 image per segmentation. 

For the 3D methods, we used the most popular segmentation methods, DenseVoxNet~\cite{yu2017automatic}, VNet~\cite{milletari2016v}, and 3D U-Net~\cite{cciccek20163d}. 

Table~\ref{internal} and Table~\ref{prostatex} compare nnU-Net based architectures on our internal dataset and ProstateX, respectively, while Table~\ref{internal++} and Table~\ref{prostatex++} compares nnU-Net++ based architectures on our internal dataset and ProstateX, respectively. Our method outperforms every other method in almost every metric. Conventional 2.5D methods are better than 2D and 3D methods for nnU-Net based methods, but the performance starts to drop as the method uses a 7-image stack as input. For nnU-Net++ based methods, 2.5D methods show little or no improvement over 2D methods and the optimal number of slices to include is unclear. Naively stacking nearby slices would not be able to fully utilize the newly added information from nearby slices without any explicit information exchange mechanism. As is shown in the tables, using either nnU-Net or nnU-Net++ as the backbone along with our CAT modules yields better performance than any of the current methods. The cross-slice attention in our method enables the network to learn the relationship between slices, which results in better performance. 
 
Four tables demonstrate that 2.5D methods were usually performing the best outside of our method. However, the optimal number of adjacent slices differed between datasets, prostate zones, and backbone networks. For example, for nnU-Net based methods, nnU-Net w/5 was the best on our internal dataset while nnU-Net w/3 was the best on ProstateX. In contrast, our method was consistently the best on different datasets, prostate zones, and backbone networks.

To perform more rigorous evaluation, we selected the best performing 2D, 2.5D and 3D methods in the nnU-Net based comparison on our internal dataset, i.e. Table~\ref{internal}, which are nnU-Net w/1, nnU-Net w/5 and 3D U-Net, and performed 5-fold cross validation to further show the robustness of our method. The results are shown in Table~\ref{best_crossval}, where the p-values were calculated based on the Mann-Whitney U Test between CAT-nnU-Net and other competing methods. The experiment further shows that our method significantly improves the performance of existing models.

\begin{table*} \setlength{\tabcolsep}{4pt} 
\caption{Comparison between nnU-Net based methods and other methods on our internal dataset.}
\centering
\begin{tabular}{ ll cccc cccc }
\toprule
&& \multicolumn{4}{c}{TZ} & \multicolumn{4}{c}{PZ} \\
\cmidrule(lr){3-6} \cmidrule(lr){7-10}
&& IoU (\%)$\uparrow$ & Dice (\%)$\uparrow$ & RAVD ($\%$)$\downarrow$ & ASSD (mm)$\downarrow$ & IoU (\%)$\uparrow$ & Dice (\%)$\uparrow$ &
RAVD ($\%$)$\downarrow$ & ASSD (mm)$\downarrow$\\
\midrule
\multicolumn{1}{l}{\multirow{5}{*}{2D}}&DeepLabV3+~\cite{chen2018encoder}& 78.4&87.8&24.2&0.285&67.4&79.9&38.1&0.479\\
&Liu et al.'s~\cite{liu2020exploring}& 77.7&87.3&25.6&0.292&69.2&81.3&35.9&0.455\\
%&U-Net++~\cite{zhou2018unet++}& 79.8&88.6&22.5&0.264&72.3&83.3&32.2&0.491\\
&CE-Net~\cite{gu2019net}&78.1&87.5&25.3&0.253&68.5&80.7&36.1&0.524\\
&MSU-Net~\cite{su2021msu}&79.8&88.6&22.4&0.214&70.7&82.3&34.1&0.431\\
&nnU-Net w/1~\cite{isensee2018nnu}& 80.0&88.7&22.4&0.215&71.9&83.4&34.2&0.347\\

\midrule
\multicolumn{1}{l}{\multirow{3}{*}{2.5D}}&nnU-Net w/3 & 80.7&89.1&22.8&0.206&72.3&83.6&33.5&0.336\\
&nnU-Net w/5 & 80.9&89.3&21.9&0.200&72.9&84.1&33.7&0.307\\
&nnU-Net w/7 & 79.6&88.5&23.2&0.222&73.4&84.4&30.7&0.291\\
\midrule
\multicolumn{1}{l}{\multirow{3}{*}{3D}}
&DenseVoxNet~\cite{yu2017automatic}& 73.5&84.4&31.1&0.343&61.4&75.4&56.2&0.623\\
&VNet~\cite{milletari2016v}& 74.7&85.2&30.7&0.311&63.7&77.5&46.2&0.547\\
&3D U-Net~\cite{cciccek20163d}& 74.6&85.2&30.3&0.312&67.1&80.0&40.6&0.418\\
\midrule
&CAT-nnU-Net & \bfseries82.6&\bfseries90.4&\bfseries19.3&\bfseries0.175&\bfseries75.8&\bfseries86.1&\bfseries28.2&\bfseries0.259\\
\bottomrule
\end{tabular}
\label{internal}
\end{table*}

\begin{table*} \setlength{\tabcolsep}{4pt}
\caption{Comparison between nnU-Net based methods and other methods on ProstateX.}
\centering
\begin{tabular}{ ll cccc cccc }
\toprule
&& \multicolumn{4}{c}{TZ} & \multicolumn{4}{c}{PZ} \\
\cmidrule(lr){3-6} \cmidrule(lr){7-10}
&& IoU (\%)$\uparrow$ & Dice (\%)$\uparrow$ & RAVD ($\%$)$\downarrow$ & ASSD (mm)$\downarrow$ & IoU (\%)$\uparrow$ & Dice (\%)$\uparrow$ &
RAVD ($\%$)$\downarrow$ & ASSD (mm)$\downarrow$\\
\midrule
\multicolumn{1}{l}{\multirow{5}{*}{2D}}&DeepLabV3+~\cite{chen2018encoder}&68.9&81.2&38.2&0.734&52.4&68.1&59.0&1.024\\
&Liu et al.'s~\cite{liu2020exploring}& 68.9&81.3&37.9&0.679&52.1&67.9&61.9&1.002\\
%&U-Net++~\cite{zhou2018unet++}& 79.8&88.6&22.5&0.264&72.3&83.3&32.2&0.491\\
&CE-Net~\cite{gu2019net}&68.0&80.4&38.4&0.849&51.2&67.1&63.9&1.013\\
&MSU-Net~\cite{su2021msu}&70.1&82.2&35.0&0.622&53.9&69.2&61.9&0.985\\
&nnU-Net w/1~\cite{isensee2018nnu}& 70.1&82.1&36.9&0.737&53.5&68.9&58.4&1.055\\
\midrule
\multicolumn{1}{l}{\multirow{3}{*}{2.5D}}&nnU-Net w/3 & 72.5&83.8&33.2&\bfseries0.583&55.8&71.0&54.1&0.876\\
&nnU-Net w/5 & 71.7&83.1&34.0&0.634&56.2&71.3&54.3&\bfseries0.864\\
&nnU-Net w/7 & 71.3&82.9&34.7&0.618&54.3&69.7&57.6&0.970\\
\midrule
\multicolumn{1}{l}{\multirow{3}{*}{3D}}
&DenseVoxNet~\cite{yu2017automatic}& 66.0&78.9&39.7&0.985&52.0&67.6&61.6&1.078\\
&VNet~\cite{milletari2016v}& 67.1&79.9&44.4&0.936&49.9&65.8&70.5&1.114\\
&3D U-Net~\cite{cciccek20163d}& 68.0&80.4&41.7&0.902&51.1&66.5&58.4&1.149\\
\midrule
&CAT-nnU-Net & \bfseries72.7&\bfseries83.9&\bfseries32.4&0.637&\bfseries58.5&\bfseries73.1&\bfseries52.5&0.890\\
\bottomrule
\end{tabular}
\label{prostatex}
\end{table*}

\begin{table*} \setlength{\tabcolsep}{4pt}
\caption{Comparison between nnU-Net++ based methods and other methods on our internal dataset.}
\centering
\begin{tabular}{ ll cccc cccc }
\toprule
&& \multicolumn{4}{c}{TZ} & \multicolumn{4}{c}{PZ} \\
\cmidrule(lr){3-6} \cmidrule(lr){7-10}
&& IoU (\%)$\uparrow$ & Dice (\%)$\uparrow$ & RAVD ($\%$)$\downarrow$ & ASSD (mm)$\downarrow$ & IoU (\%)$\uparrow$ & Dice (\%)$\uparrow$ &
RAVD ($\%$)$\downarrow$ & ASSD (mm)$\downarrow$\\
\midrule
\multicolumn{1}{l}{\multirow{5}{*}{2D}}&DeepLabV3+~\cite{chen2018encoder}& 78.4&87.8&24.2&0.285&67.4&79.9&38.1&0.479\\
&Liu et al.'s~\cite{liu2020exploring}& 77.7&87.3&25.6&0.292&69.2&81.3&35.9&0.455\\
%&U-Net++~\cite{zhou2018unet++}& 79.8&88.6&22.5&0.264&72.3&83.3&32.2&0.491\\
&CE-Net~\cite{gu2019net}&78.1&87.5&25.3&0.253&68.5&80.7&36.1&0.524\\
&MSU-Net~\cite{su2021msu}&79.8&88.6&22.4&0.214&70.7&82.3&34.1&0.431\\
&nnU-Net++ w/1~\cite{zhou2018unet++}& 82.2&\bfseries90.1&20.5&0.192&75.7&85.9&28.7&0.268\\
\midrule
\multicolumn{1}{l}{\multirow{3}{*}{2.5D}}
&nnU-Net++ w/3 & 81.8&89.8&20.3&0.190&74.8&85.4&30.1&0.265\\
&nnU-Net++ w/5 & 81.5&89.7&21.2&0.291&74.9&85.4&29.8&0.287\\
&nnU-Net++ w/7 & 81.6&89.8&20.8&0.192&74.5&85.1&28.7&0.269\\
\midrule
\multicolumn{1}{l}{\multirow{3}{*}{3D}}
&DenseVoxNet~\cite{yu2017automatic}& 73.5&84.4&31.1&0.343&61.4&75.4&56.2&0.623\\
&VNet~\cite{milletari2016v}& 74.7&85.2&30.7&0.311&63.7&77.5&46.2&0.547\\
&3D U-Net~\cite{cciccek20163d}& 74.6&85.2&30.3&0.312&67.1&80.0&40.6&0.418\\
\midrule
&CAT-nnU-Net++ & \bfseries82.3&\bfseries90.1&\bfseries20.0&\bfseries0.188&\bfseries76.1&\bfseries86.3&\bfseries28.3&\bfseries0.255\\
\bottomrule
\end{tabular}
\label{internal++}
\end{table*}

\begin{table*} \setlength{\tabcolsep}{4pt}
\caption{Comparison between nnU-Net++ based methods and other methods on ProstateX.}
\centering
\begin{tabular}{ ll cccc cccc }
\toprule
&& \multicolumn{4}{c}{TZ} & \multicolumn{4}{c}{PZ} \\
\cmidrule(lr){3-6} \cmidrule(lr){7-10}
&& IoU (\%)$\uparrow$ & Dice (\%)$\uparrow$ & RAVD ($\%$)$\downarrow$ & ASSD (mm)$\downarrow$ & IoU (\%)$\uparrow$ & Dice (\%)$\uparrow$ &
RAVD ($\%$)$\downarrow$ & ASSD (mm)$\downarrow$\\
\midrule
\multicolumn{1}{l}{\multirow{5}{*}{2D}}&DeepLabV3+~\cite{chen2018encoder}&68.9&81.2&38.2&0.734&52.4&68.1&59.0&1.024\\
&Liu et al.'s~\cite{liu2020exploring}& 68.9&81.3&37.9&0.679&52.1&67.9&61.9&1.002\\
%&U-Net++~\cite{zhou2018unet++}& 79.8&88.6&22.5&0.264&72.3&83.3&32.2&0.491\\
&CE-Net~\cite{gu2019net}&68.0&80.4&38.4&0.849&51.2&67.1&63.9&1.013\\
&MSU-Net~\cite{su2021msu}&70.1&82.2&35.0&0.622&53.9&69.2&61.9&0.985\\
&nnU-Net++ w/1~\cite{zhou2018unet++}& 72.0&83.5&32.9&0.594&56.1&71.1&53.3&0.865\\
\midrule
\multicolumn{1}{l}{\multirow{3}{*}{2.5D}}
&nnU-Net++ w/3 & 72.8&84.0&32.4&0.535&56.5&71.6&55.1&0.847\\
&nnU-Net++ w/5 & 71.0&82.7&33.8&0.656&56.7&71.8&53.5&0.855\\
&nnU-Net++ w/7 & 71.9&83.3&33.0&0.615&56.8&71.8&57.2&0.874\\
\midrule
\multicolumn{1}{l}{\multirow{3}{*}{3D}}
&DenseVoxNet~\cite{yu2017automatic}& 66.0&78.9&39.7&0.985&52.0&67.6&61.6&1.078\\
&VNet~\cite{milletari2016v}& 67.1&79.9&44.4&0.936&49.9&65.8&70.5&1.114\\
&3D U-Net~\cite{cciccek20163d}& 68.0&80.4&41.7&0.902&51.1&66.5&58.4&1.149\\
\midrule
&CAT-nnU-Net++ & \bfseries73.1&\bfseries84.1&\bfseries29.6&\bfseries0.512&\bfseries58.2&\bfseries73.1&\bfseries52.3&\bfseries0.781\\
\bottomrule
\end{tabular}
\label{prostatex++}
\end{table*}

\begin{table*} \setlength{\tabcolsep}{4pt} 
\caption{Comparison between nnU-Net based methods and other methods on our internal dataset, where * indicates p-value $\leq0.1$, ** indicates p-value less than $\leq0.05$ and *** indicates p-value less than $\leq0.01$.}
\centering
\begin{tabular}{ l cccc cccc }
\toprule
& \multicolumn{4}{c}{TZ} & \multicolumn{4}{c}{PZ} \\
\midrule
& IoU (\%)$\uparrow$ & Dice (\%)$\uparrow$ & RAVD ($\%$)$\downarrow$ & ASSD (mm)$\downarrow$ & IoU (\%)$\uparrow$ & Dice (\%)$\uparrow$ &
RAVD ($\%$)$\downarrow$ & ASSD (mm)$\downarrow$\\
\midrule
nnU-Net w/1~\cite{isensee2018nnu}& 79.6***&88.5***&23.6***&0.243***&72.2***&83.6***&32.8***&0.371**\\
nnU-Net w/5 & 80.7***&89.2***&22.3 ***&0.204***&72.7***&84.0***&32.5***&0.354**\\
3D U-Net~\cite{cciccek20163d}&75.6***&85.8***&29.1***&0.301***&66.6***&79.6***&41.2***&0.430***\\
CAT-nnU-Net & \bfseries81.3&\bfseries89.5&\bfseries20.8&\bfseries0.201&\bfseries74.3&\bfseries85.1&\bfseries29.8&\bfseries0.301\\
\bottomrule
\end{tabular}
\label{best_crossval}
\end{table*}

\begin{table*} \setlength{\tabcolsep}{4pt}
\caption{Ablation study on the effect of nn-based design and CAT modules on U-Net.}
\centering
\begin{tabular}{ cc cccc cccc }
\toprule
\multirow{3}{*}{\thead[l]{nn}} &
\multirow{3}{*}{\thead[l]{CAT}} &
\multicolumn{4}{c}{TZ} & \multicolumn{4}{c}{PZ} \\
\cmidrule(lr){3-6} \cmidrule(lr){7-10}
&& IoU (\%)$\uparrow$ & Dice (\%)$\uparrow$ & RAVD ($\%$)$\downarrow$ & ASSD (mm)$\downarrow$ & IoU (\%)$\uparrow$ & Dice (\%)$\uparrow$ &
RAVD ($\%$)$\downarrow$ & ASSD (mm)$\downarrow$\\
\midrule
& & 77.5&87.2&25.7&0.311&68.8&80.8&36.1&0.665\\
\checkmark& & 80.0&88.7&22.4&0.215&71.9&83.4&34.2&0.347\\
&
\checkmark&75.4&85.4&26.3&0.319&70.5&82.1&33.0&0.397\\
\checkmark& \checkmark& \bfseries82.6&\bfseries90.4&\bfseries19.3&\bfseries0.175&\bfseries75.8&\bfseries86.1&\bfseries28.2&\bfseries0.259\\
\bottomrule
\end{tabular}
\label{ablation_unet}
\end{table*}

\begin{table*} \setlength{\tabcolsep}{4pt}
\caption{Ablation study on the effect of nn-based design and CAT modules on U-Net++.}
\centering
\begin{tabular}{ cc cccc cccc }
\toprule
\multirow{3}{*}{\thead[l]{nn}} &
\multirow{3}{*}{\thead[l]{CAT}} &
\multicolumn{4}{c}{TZ} & \multicolumn{4}{c}{PZ} \\
\cmidrule(lr){3-6} \cmidrule(lr){7-10}
&& IoU (\%)$\uparrow$ & Dice (\%)$\uparrow$ & RAVD ($\%$)$\downarrow$ & ASSD (mm)$\downarrow$ & IoU (\%)$\uparrow$ & Dice (\%)$\uparrow$ &
RAVD ($\%$)$\downarrow$ & ASSD (mm)$\downarrow$\\
\midrule
& & 77.9&87.4&24.7&0.313&69.0&81.0&35.0&0.507\\
\checkmark& & 82.2&\bfseries90.1&20.5&0.192&75.7&85.9&28.7&0.268\\
&
\checkmark&78.8&87.9&23.8&0.256&70.9&82.4&34.4&0.419\\
\checkmark& \checkmark& \bfseries82.3&\bfseries90.1&\bfseries20.0&\bfseries0.188&\bfseries76.1&\bfseries86.3&\bfseries28.3&\bfseries0.255\\
\bottomrule
\end{tabular}
\label{ablation_unetplusplus}
\end{table*}

\begin{table*} \setlength{\tabcolsep}{4pt}
\caption{Ablation study on the effect of positional encoding and transformer blocks, where * indicates p-value $\leq0.1$, ** indicates p-value less than $\leq0.05$ and *** indicates p-value less than $\leq0.01$.}
\centering
\begin{tabular}{ cc cccc cccc }
\toprule
\multirow{3}{*}{\thead[l]{Positional\\Encoding}} &
\multirow{3}{*}{\thead[l]{Transformer\\Blocks}} &
\multicolumn{4}{c}{TZ} & \multicolumn{4}{c}{PZ} \\
\cmidrule(lr){3-6} \cmidrule(lr){7-10}
&& IoU (\%)$\uparrow$ & Dice (\%)$\uparrow$ & RAVD ($\%$)$\downarrow$ & ASSD (mm)$\downarrow$ & IoU (\%)$\uparrow$ & Dice (\%)$\uparrow$ &
RAVD ($\%$)$\downarrow$ & ASSD (mm)$\downarrow$\\
\midrule
& & 79.6***&88.5**&23.6 ***&0.243***&72.2***&83.6***&32.8***&0.371**\\
\checkmark& & 80.5**&89.0*&23.0**&0.220*&73.1***&84.2***&32.3***&0.350**\\
&
\checkmark&80.6*&89.1*&21.8**&0.202 ***&\bfseries74.4&\bfseries85.1&30.7&0.302\\
\checkmark& \checkmark& \bfseries81.3&\bfseries89.5&\bfseries20.8&\bfseries0.201&74.3&\bfseries85.1&\bfseries29.8&\bfseries0.301\\
\bottomrule
\end{tabular}
\label{ablation}
\end{table*}

\begin{comment}
\begin{table*} \setlength{\tabcolsep}{4pt}
\caption{Ablation study on the effect of positional encoding and transformer blocks.}
\centering
\begin{tabular}{ cc cccc cccc }
\toprule
\multirow{3}{*}{\thead[l]{Positional\\Encoding}} &
\multirow{3}{*}{\thead[l]{Transformer\\Blocks}} &
\multicolumn{4}{c}{TZ} & \multicolumn{4}{c}{PZ} \\
\cmidrule(lr){3-6} \cmidrule(lr){7-10}
&& IoU (\%)$\uparrow$ & Dice (\%)$\uparrow$ & RAVD ($\%$)$\downarrow$ & ASSD (mm)$\downarrow$ & IoU (\%)$\uparrow$ & Dice (\%)$\uparrow$ &
RAVD ($\%$)$\downarrow$ & ASSD (mm)$\downarrow$\\
\midrule
& & 80.0&88.7&22.4&0.215&71.9&83.4&34.2&0.347\\
\checkmark& & 82.1&89.9&20.1&0.206&75.2&85.5&28.9&0.266\\
&
\checkmark&\bfseries82.6&\bfseries90.4&19.4&0.176&74.9&85.4&28.9&0.266\\
\checkmark& \checkmark& \bfseries82.6&\bfseries90.4&\bfseries19.3&\bfseries0.175&\bfseries75.8&\bfseries86.1&\bfseries28.2&\bfseries0.259\\
\bottomrule
\end{tabular}
\label{ablation}
\end{table*}
\end{comment}
 
\begin{table*} \setlength{\tabcolsep}{4pt}
\caption{Comparison between using CAT modules in all layers vs only in Layers 0--3.}
    \centering
\begin{tabular}{ l cccc cccc }
\toprule
& \multicolumn{4}{c}{TZ} & \multicolumn{4}{c}{PZ} \\
\cmidrule(lr){2-5} \cmidrule(lr){6-9}
& IoU (\%)$\uparrow$ & Dice (\%)$\uparrow$ & RAVD ($\%$)$\downarrow$ & ASSD (mm)$\downarrow$ & IoU (\%)$\uparrow$ & Dice (\%)$\uparrow$ &
RAVD ($\%$)$\downarrow$ & ASSD (mm)$\downarrow$\\
\midrule
Without CAT modules in Layers 4 and 5& 82.4&90.2&19.7&0.188&\bfseries76.3&\bfseries86.4&\bfseries27.4&\bfseries0.244\\
With CAT modules in Layers 4 and 5&  \bfseries82.6&\bfseries90.4&\bfseries19.3&\bfseries0.175&75.8&86.1&28.2&0.259\\
\hline
\end{tabular}
\label{attention}
\end{table*}

\subsubsection{Ablation Study}
\label{sec_ablation}
To support the claim that nnU-Net based network architectures have better performance than conventional U-Net based network architectures, we conducted the following ablation studies, where we compared the performance of using CAT modules or not on either nn or conventional architectured U-Net and U-Net++ with our internal dataset. The results are shown in Table~\ref{ablation_unet} and Table~\ref{ablation_unetplusplus}, where CAT modules improved the the segmentation of PZ by a lot on conventional network architectures. In general, nn-based networks outperform conventional networks, while applying CAT modules on nn-based networks yields the best results.

Due to the larger improvement in performance when adding CAT modules to nnU-Net and the long training time of CAT-nnU-Net++, we performed our ablation study on the effect of positional encoding and transformer blocks using only the CAT-nnU-Net. The results on our internal dataset are reported in Table~\ref{ablation}, where we used 5-fold cross validation and the p-values were calculated based on Mann-Whitney U Test between the network using all the components and the other networks missing positional encoding or transformer blocks. From the table we can conclude that using both positional encoding and Transformer blocks can help with prostate zonal segmentation. Using both or either one alone outperforms using neither. Using transformer blocks alone would have comparable results to using both in PZ segmentation as there is no statistical significance, but using both significantly outperforms only using transformer blocks. Besides, using both is also better than just using positional encoding.

 \begin{figure}
     \centering
     \subfloat[]{
         \includegraphics[height=0.325\textwidth]{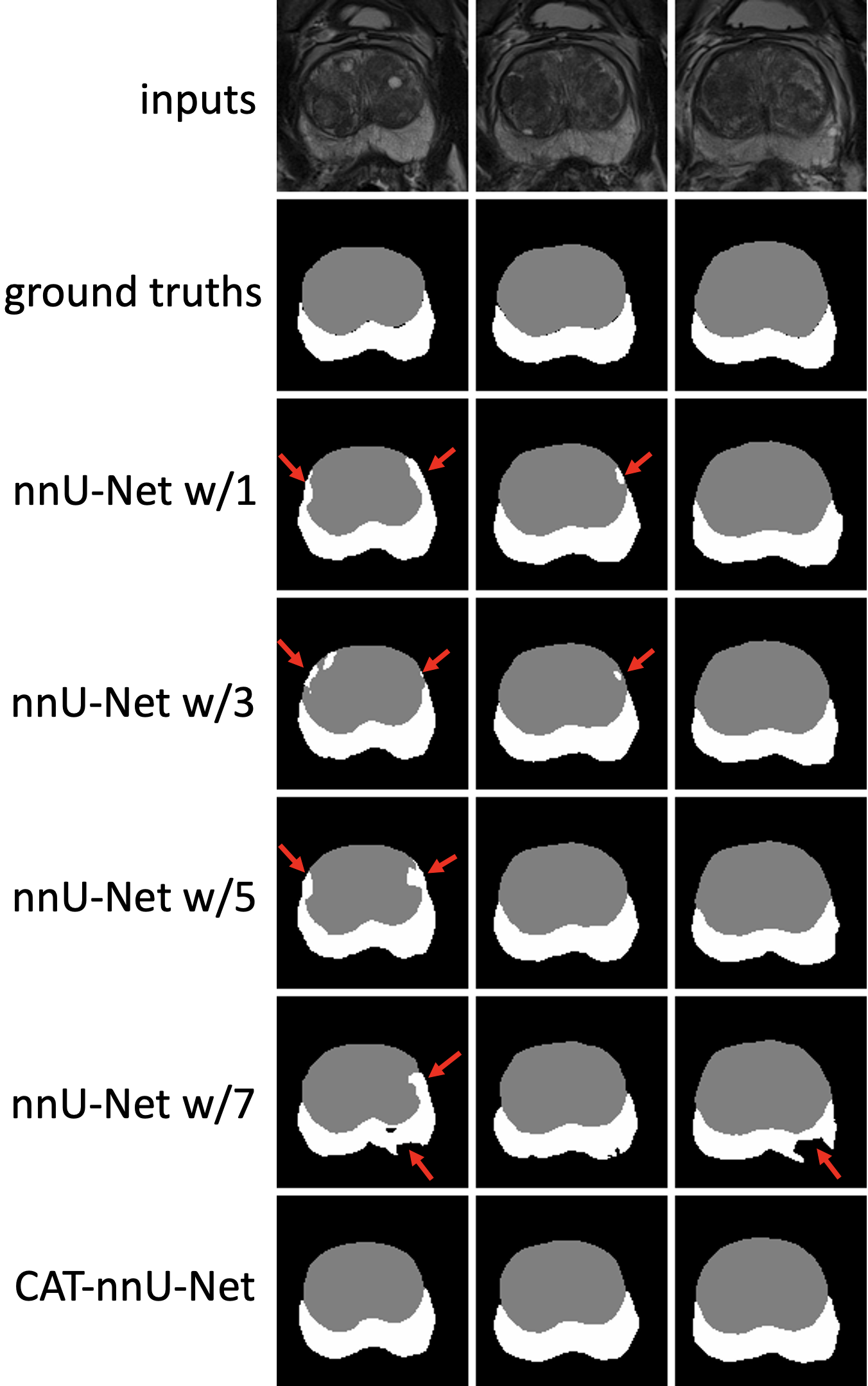}
         \label{nnu}
         }
         \subfloat[]{
         \includegraphics[height=0.325\textwidth]{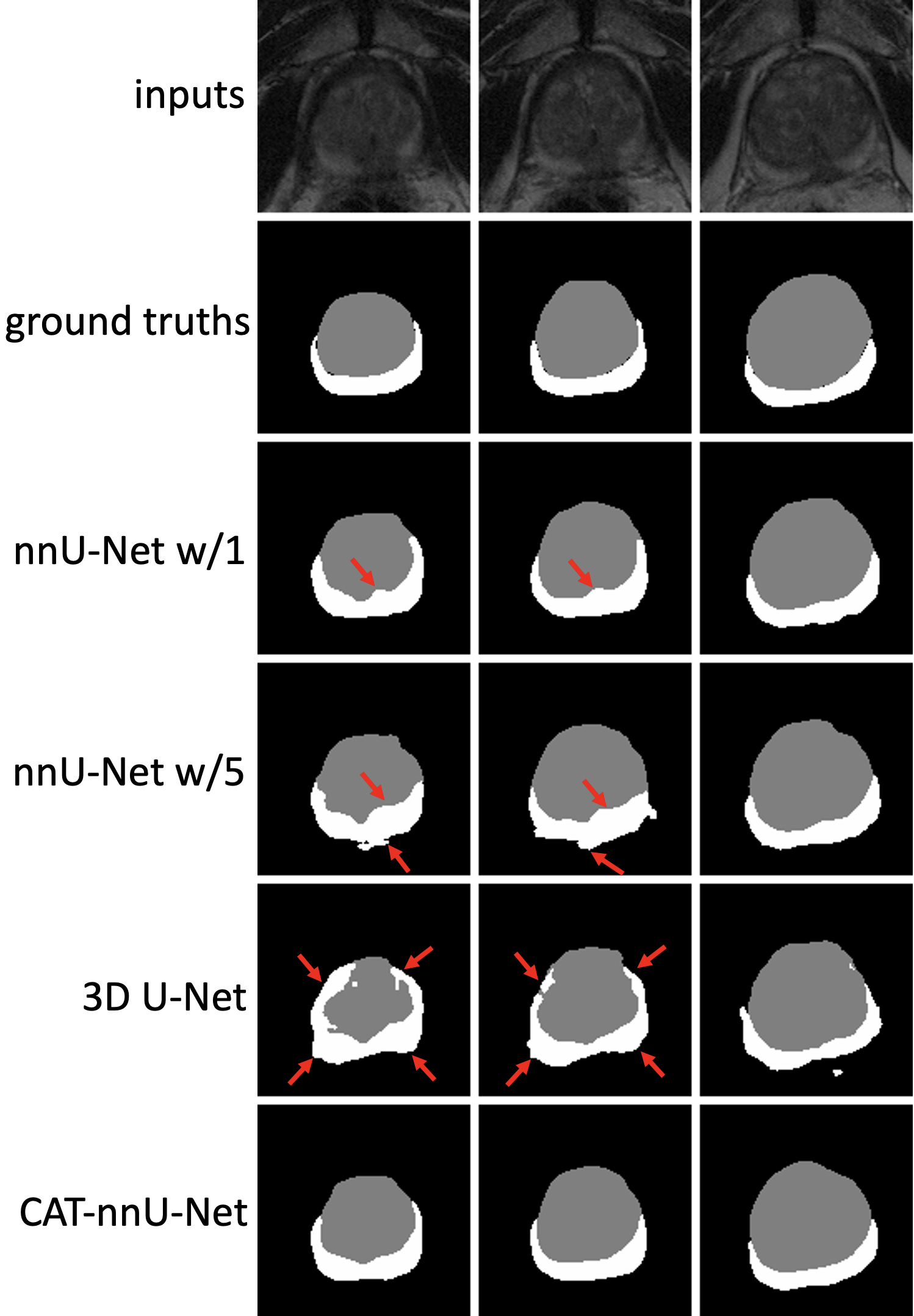}
         \label{best}
         }
     \caption{Comparison of the CAT-nnU-Net against (a) other nnU-Net based methods and (b) nnU-Net w/1, nnU-Net w/5, and 3D U-Net. The segmentation masks of PZ are shown in white, those of TZ in gray, and those of other tissues in black.}
     \label{qualitative}
 \end{figure}
 
 \begin{figure*}
     \centering
     \subfloat[]{
         \includegraphics[width=0.485\textwidth]{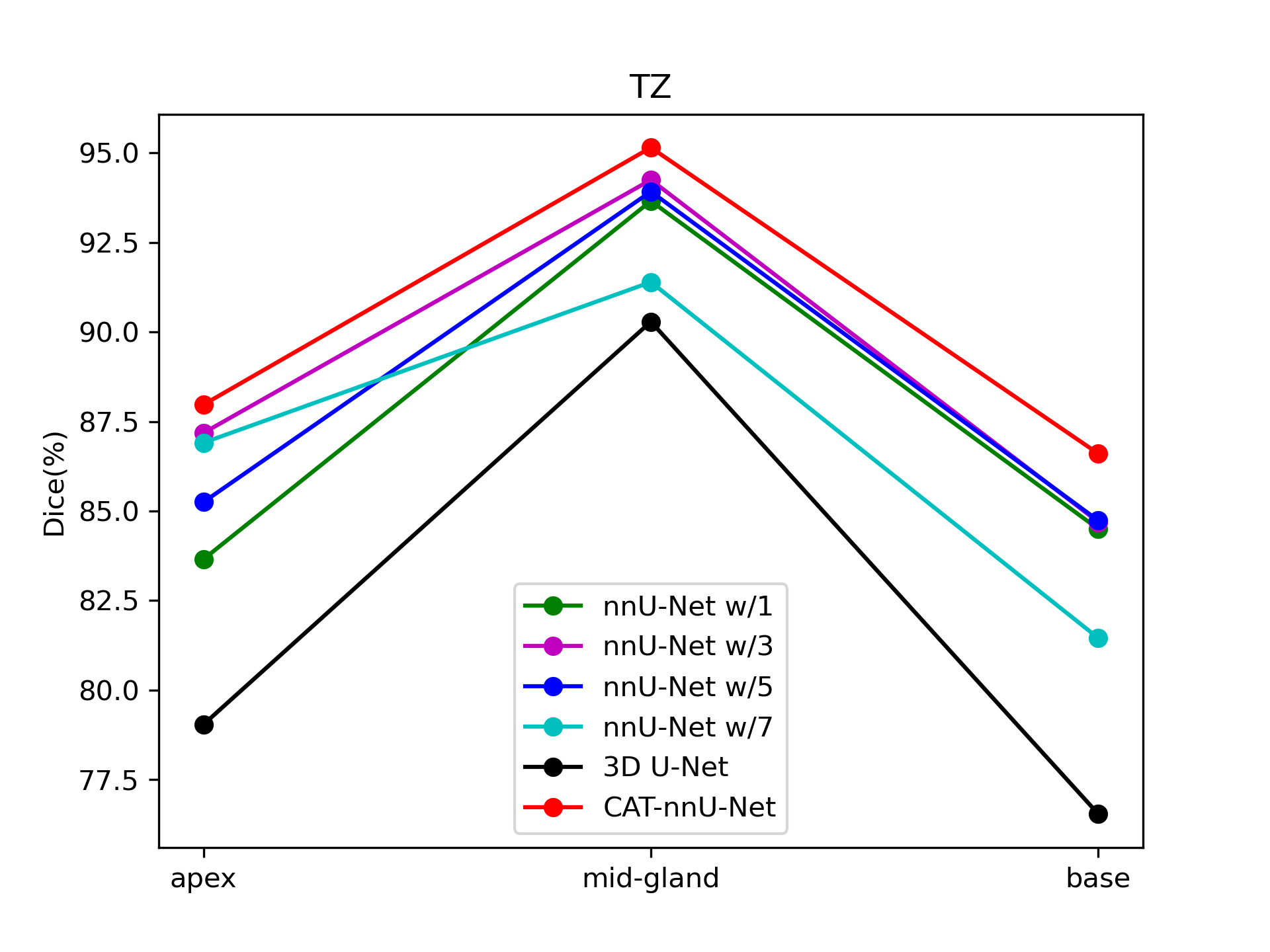}
         \label{tz}
         }
         \subfloat[]{
         \includegraphics[width=0.485\textwidth]{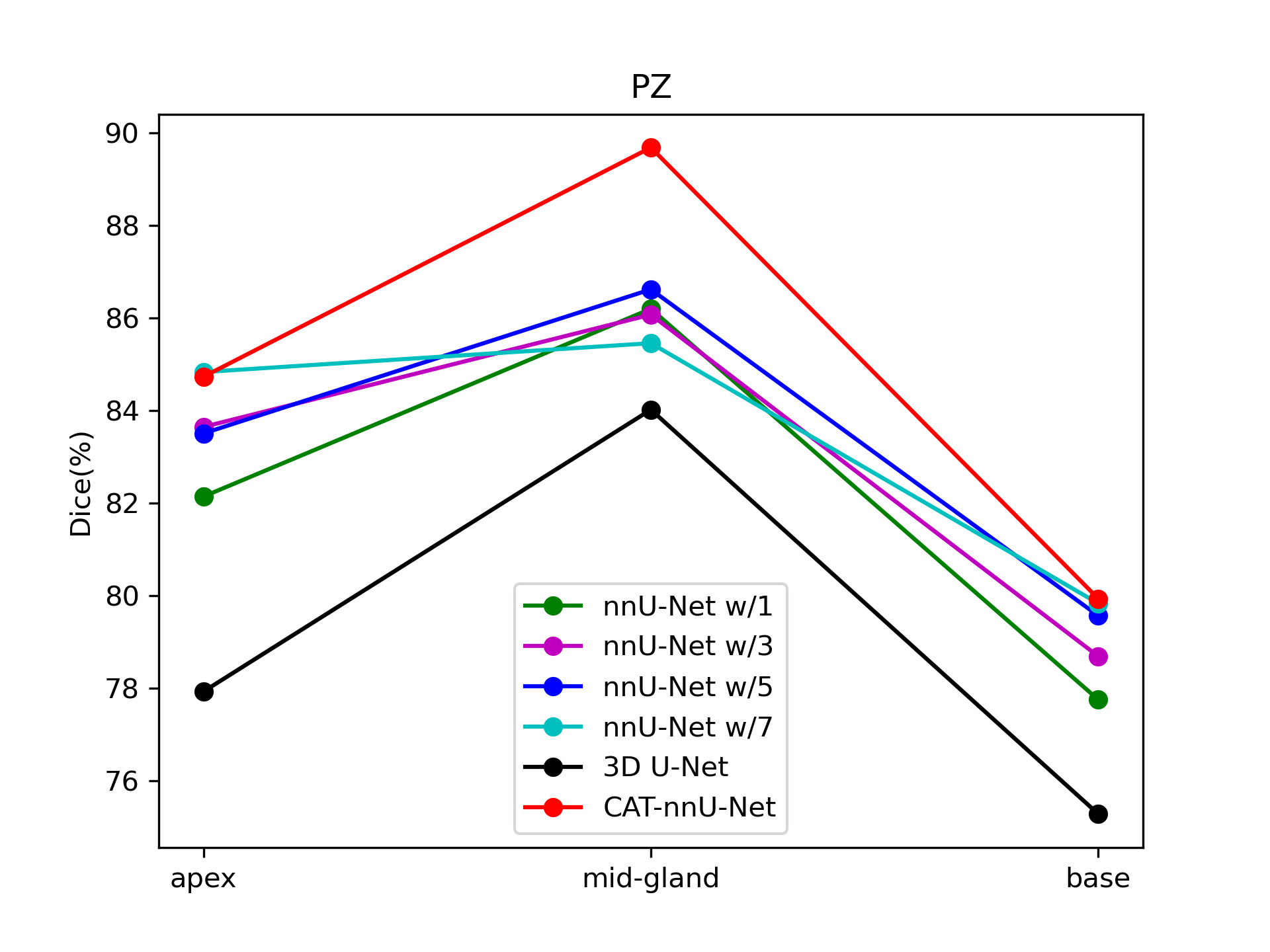}
         \label{pz}
         }
     \caption{The performance of (a) TZ and (b) PZ segmentation on different prostate parts by different algorithms.}
     \label{region}
 \end{figure*}

\subsubsection{Qualitative Evaluation}

Some representative results on our internal dataset are shown in Fig.~\ref{qualitative}. In Fig.~\ref{nnu}, we compare our CAT-nnU-Net with other nnU-Net based 2D and 2.5D methods. Other nnU-Net based methods tend to overpredict PZ on the side of the TZ. 
%The left example in Fig.~\ref{nnu} shows three consecutive prostate T2WI slices from a patient with suspicious PCa in the right posterior side of PZ, which is located in the lower left of the T2WI images shown as a small dark area surrounded by brighter regions. As we can observe, since the intensity of region of the suspicious lesion is different compared with the intensity of surrounded PZ, while being more similar to TZ, all models except ours mis-identify the region around the suspicious lesion as part of TZ. Our model has demonstrated better robustness in the segmentation of PZ and TZ even when there are suspicious lesions in the images. In the right example in Fig.~\ref{nnu}, similar to what we have described in Fig.~\ref{best}, the other models perform badly on the segmentation of PZ especially on the two upper head regions. We believe this is due to their lack of ability to accurately capture the cross-slice relationship. In comparison, our method can perform accurate segmentation around the upper head region of PZ that the other methods fail to generate good prediction masks.
In the example of Fig.~\ref{nnu}, our model performs better in the segmentation of PZ, especially in the lower middle region and the two upper head regions of the PZ. Without effectively considering the cross-slice relationship, e.g., where the PZ start to emerge, the models would poorly segment the upper head region of the PZ. This phenomenon is clearly revealed by the results of the other methods, which over-predict the PZ region with many false positive predictions. Additionally, when including 7 slices in the 2.5D segmentation, the segmentation of the lower part of the PZ produces spurious spur-like regions. As the red arrows indicate in Fig.~\ref{nnu}, there are obvious errors in the segmentation, but the segmentation produced by our method does not suffer from such errors.

%In Fig.~\ref{best}, for a clearer demonstration of the superiority of our method, we compared nnU-Net + ISIMs against the best performing 2D, 2.5D nnU-Net based methods and the best 3D method based on the quantitative results, which is nnU-Net w/1, nnU-Net w/5 and 3D U-Net. The left example in Fig.~\ref{best} shows that our proposed model performs better in the segmentation of PZ, especially in the lower middle region and the two upper head regions of the PZ. Without effectively consider the cross-slice relationship, e.g. where the PZ start to emerge, the models would be confused about the segmentation of the upper head region of the PZ. This phenomenon is clearly shown in the results of other methods, which over-predict PZ region with many false positive predictions. 

For a clearer demonstration of CAT-nnU-Net's superiority, in Fig.~\ref{best} we compare it against the best performing 2D, 2.5D nnU-Net based methods and the best 3D method based on the quantitative results, which are nnU-Net w/1, nnU-Net w/5, and 3D U-Net. Consistent with Fig.~\ref{nnu}, the other models perform badly on the segmentation of PZ especially on the two upper head regions, due to their inability to accurately capture the cross-slice relationship. By contrast, our method can perform accurate segmentation around the upper head region of the PZ where the other methods fail to generate satisfactory prediction masks. Additionally, due to the poor imaging quality, the other methods fail to accurately segment the boundaries of the PZ, as are indicated by the red arrows.

 \begin{figure*}
     \centering{
         \includegraphics[width=0.85\textwidth]{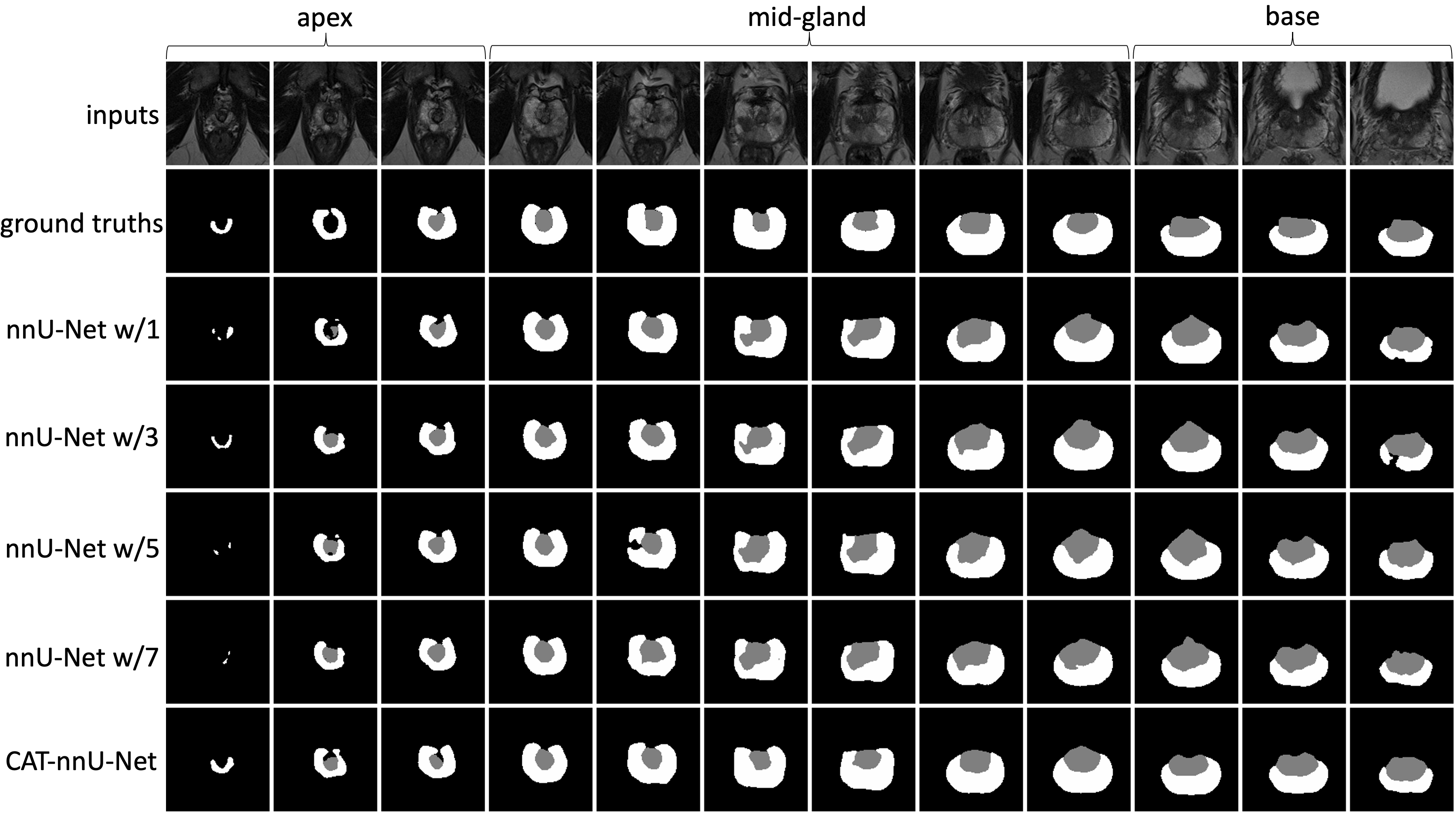}
         }
     \caption{Comparison of our CAT-nnU-Net against other nnU-Net based methods on different prostate parts.}
     \label{apexbase}
 \end{figure*}
 
 \subsubsection{Evaluation on Different Parts of the Prostate}
\label{different_regions}

 Prostate zonal segmentation approaches exhibit significant performance differences in different parts of the prostate~\cite{liu2020exploring}; hence, we will next investigate the performance of our method on different parts of the prostate on our internal dataset. We define the first 3 and last 3 slices of the prostate as apex and base, respectively, and the remaining slices are mid-gland slices. We continue our evaluation against other nnU-Net based methods and 3D U-Net. As shown in Fig.~\ref{region}, other nnU-Net based methods have different performance in different parts of the prostate for different zones. Among the other nnU-Net based methods, nnU-Net w/5 is the best for TZ segmentation in all parts based on the quantitative results, while it performs badly in apex slices. Although nnU-Net w/7 performs well in segmenting PZ in the apex and base slices, its performance in the mid-gland slices and TZ segmentation is underwhelming. Therefore, it would be hard to determine how many slices to include when performing traditional 2.5D prostate zonal segmentation. However, our method inputs all the slices and learns the relationship between slices. Although our method yields a marginally worse result than nnU-Net w/7 on apex for PZ, it clearly outperforms other nnU-Net based methods in general.

Fig.~\ref{apexbase} shows qualitative results of prostate zonal segmentation on a single patient. Among the other nnU-Net based methods, our CAT-nnU-Net is the most consistent, whereas other methods suffer from over-prediction and inconsistency in anatomical structures.

\subsubsection{Understanding the Attention Matrices}

\begin{figure}
     \centering
         \includegraphics[width=0.48\textwidth]{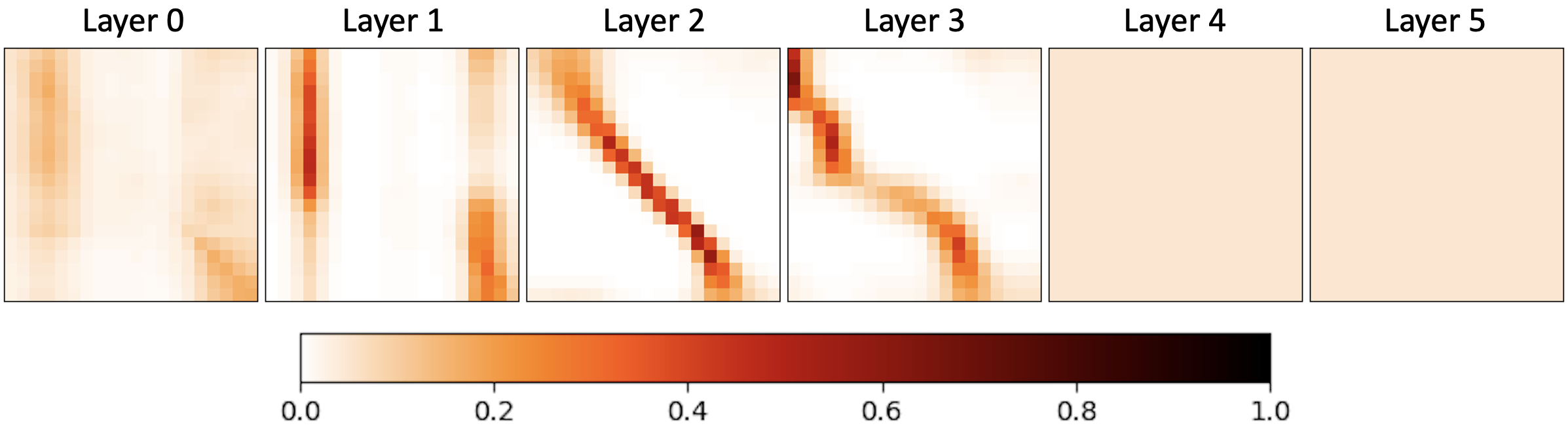}

     \caption{Attention matrices $A$ from Layers 0--5. Each element $A[i,j]$ indicates how much attention slice $i$ pays to slice $j$, with darker pixels indicating higher values, and $\sum_{j}A[i,j]=1$.}
     \label{attention_vis}
 \end{figure}

Since CAT-nnU-Net yields a larger improvement over nnU-Net than CAT-nnU-Net++ yields over nnU-Net++, we used the former in the following experiment. The most notable attention matrices $A$ from each layer are visualized in Fig.~\ref{attention_vis}, ranging from fine-resolution, top Layer~0 to coarse-resolution, bottom Layer~5. The CAT modules in the finer Layers~0--3 were able to learn something meaningful, while those in the coarser Layers~4 and 5 learned nothing. This observation is supported by the segmentation results reported in Table~\ref{attention}, which are similar with and without attention in Layers~4 and 5. The full network has slightly better results on TZ but it has worse results on PZ. The cross-slice attention in the coarsest layers may confuse the network when segmenting the more challenging PZ. 

Apparently, the cross-slice attention mechanism is unable to learn anything useful in Layers~4 and 5 because the network need not rely on nearby slices to learn coarse information. This seems to be consistent with how clinicians annotate prostate images, as they need to refer to nearby slices only to segment the finer details. Though the CAT modules in Layers~4 and 5 learn nothing useful, we include them nonetheless, since doing so does not significantly hurt performance. The attention matrices in Layers~0 and 1 indicate that for the details, mid-gland slices will attend more to either the apex or the base slices, and slices in the apex and base will attend more to the slices in their own prostate part. At the same time, apex and base slices will also attend a little bit to each other, which shows that there are some long-range dependencies in the segmentation of finer details. The attention in Layers~2 and 3, which contain more significant information than Layers~0 and 1, focuses more on nearby slices in the mid-gland slices. This is in line with the results of the previous section, where more slices are needed to accurately segment the base and apex, hence the more blurry regions at the top left and bottom right of Layer~2 as well as the bottom right of Layer~3.

\section{Discussion}

Our study demonstrated that applying the CAT modules on skip connections at different scales can improve the performance of prostate zonal segmentation by systematically exploiting cross-slice attention. In particular, we see a significant improvement in PZ segmentation compared with TZ segmentation. We believe that this may be because it is relatively easier to segment TZ without accounting for nearby slices as the shape of the TZ is well-defined and consistent across subjects. By contrast, PZ segmentation is harder than TZ segmentation since the shape of the PZ is less clear in certain slices, which may be why our cross-slice attention method improved the performance of PZ segmentation relative to existing networks.

Our cross-slice attention method has shown its superiority to other methods for the following reasons: 2D networks cannot acquire useful information from nearby slices. 3D convolution approaches do not perform well due to the anisotropy of the MRI data (i.e., through-plane resolution substantially lower than in-plane resolution), which is problematic in learning the 3D shape. The problem with conventional 2.5D methods may be that the network cannot sufficiently process which slice is the one to be segmented. Additionally, including more slices in the stack introduces more useless information, so the network could have a harder time learning the useful information. Furthermore, inputting more slices might make the network more susceptible to overfitting.

We find that CAT modules yield the biggest improvement on skip-connection-based encoder-decoder networks, e.g., U-Net and U-Net++ using leaky ReLU as activation and instance normalization as normalization, which is the scheme in nnU-Net. However, even with different activation functions and normalization schemes, performance drops when applying the cross-slice attention on frameworks without skip connections, e.g., DeepLabV3+ and the method of Liu et al. The networks can learn different attention at different resolutions since the information in the feature maps at different scales differs. This enables the network to learn more meaningful cross-slice information. On the other hand, the network could be confused by the cross-slice information without skip connections at different scales, leading to worse performance. Further investigation into why the CAT modules do not work on frameworks without skip connections and how to make similar ideas work on these architectures would be a good direction for further study.  

In this work, we only considered T2WI, but in practice, as suggested by previous works~\cite{rundo2018fully}~\cite{nai2020evaluation}, performing prostate segmentation on different MRI or multispectral MRI can be considered to  improve segmentation performance. Other image contrasts, such as T1WI, are capable of providing information related to the segmentation that is not present in T2WI~\cite{ozer2010supervised}. To perform multispectral MRI tasks with our proposed framework, we can add the multispectral MRI images as different input channels to the network, where each channel is one type of MRI image. However, further consideration may be needed for zonal segmentation since T1WI generally does not contain good tissue contrast between prostate zones. 

We evaluated our method on two datasets: our internal dataset and ProstateX dataset, where we see better performance of our internal dataset since it has better annotations and thinner slices. However, the CAT module improves the performance on both datasets. Although our method was evaluated only on prostate zonal segmentation due to the limitation of the data availability, it is not limited to such a problem, and it promises to be applicable in any 3D segmentation task where information across all slices is needed. This is most useful in anisotropic image data where cross-slice information is crucial to accurate segmentation and 3D CNN methods, which are designed for isotropic image data, tend to fail. For isotropic data, our method should produce decent results, but its downside will be high memory requirements.
Note that our method is less memory efficient than those lacking CAT modules. For instance, for inputs with size of $128\times128\times20$, CAT-nnU-Net has $6.1\times10^8$ trainable parameters compared with $1.4\times10^8$ for nnU-Net, and CAT-nnU-Net++ has $3.9\times10^8$ trainable parameters as opposed to $3.7\times10^7$ for U-Net++. How to adapt our method to other problems other than prostate zonal segmentation and how to make it more memory efficient would be interesting directions for future works. 

\section{Conclusions}

 We have demonstrated improved prostate zonal segmentation by applying a self-attention mechanism to systematically learn and leverage cross-slice information. More specifically, we have proposed a CAT module and incorporated it into state-of-the-art 2D skip connection-based deep networks. Our resulting CAT-Net models exhibit better performance than the current state-of-the-art 2D, 2.5D, and 3D competitors in the task of prostate zonal segmentation, especially in the peripheral zone of the prostate. Compared with conventional 2.5D prostate segmentation methods, our segmentation performance was consistently good across all three prostate parts (apex, mid-gland, and base). Furthermore, our analysis of the attention matrices provided insights into how the cross-slice attention mechanism helps with the segmentation. Our approach has proven to be useful in networks with skip connections at different scales, and our ablation study has shown that each of its components contributes such that their combination yields the best results. Further research is needed on how best to incorporate our CAT modules into other network architectures.

\balance
\bibliographystyle{IEEEtran}
\bibliography{ref.bib}

\end{document}